\documentclass[a4paper,12pt]{article}

\usepackage{graphicx}
\usepackage{subcaption}
\usepackage{epsfig}
\usepackage{epsf}

\usepackage{amsmath}
\usepackage{amssymb}
\usepackage{amsfonts}
\usepackage{adjustbox}

\usepackage{bm}
\usepackage{float}
\usepackage{hyperref}
\usepackage{authblk}

\usepackage[normalem]{ulem}
\usepackage{xcolor}
\usepackage{cancel}

\newcommand{\jk}[1]{\textcolor{blue}{#1}}

\title{\bf
Nonlinear electroweak sphaleron clouds \\ around Kerr black holes
}

\author[1,3]{Carlos Herdeiro\thanks{\href{mailto:herdeiro@ua.pt}{herdeiro@ua.pt}}} 
\author[2]{Jutta Kunz\thanks{\href{mailto:jutta.kunz@uni-oldenburg.de}{jutta.kunz@uni-oldenburg.de}}} 
\author[2]{Hendrik Mennenga\thanks{\href{mailto:handrik.mennenga@uni-oldenburg.de}{hendrik.mennenga@uni-oldenburg.de}}} 
\author[1]{Eugen Radu\thanks{\href{mailto:eugen.radu@ua.pt}{eugen.radu@ua.pt}}}

\affil[1]{\small Departamento de Matem\'atica da Universidade de Aveiro and
Centre for Research and Development in Mathematics and Applications (CIDMA),
Campus de Santiago, 3810-183 Aveiro, Portugal}
\affil[2]{\small Institut f\"ur  Physik, Universit\"at Oldenburg, Postfach 2503,
D-26111 Oldenburg, Germany}
\affil[3]{\small Programa de P\'os-Gradua\c{c}\~{a}o em F\'{\i}sica, U. Federal do Par\'a, 66075-110, Bel\'em, Par\'a, Brazil}

\date{September 2026}

\begin{document}

\maketitle

\begin{abstract} 
Kerr black holes and electroweak sphalerons are remarkable solutions of two pillars of modern physics: General Relativity and the Standard Model of particle physics, respectively. In this work, we investigate their interplay by constructing nonlinear sphaleron clouds around Schwarzschild and Kerr black holes in the probe limit. We first revisit static and spinning (multi-)sphalerons in flat spacetime, using the measured electroweak parameters. We then explore the domain of existence and physical properties of the corresponding black-hole configurations, revealing a rich branch structure. Remarkably, the sphaleron clouds retain a finite mass gap and do not approach linear clouds within the families studied, distinguishing them from the familiar clouds associated with the onset of superradiant instabilities. We also obtain a closed-form leading-order Higgs solution on an exact Yang–Mills configuration in a Schwarzschild background.
\end{abstract}

 \clearpage

\tableofcontents
 
 \clearpage
 
\section{Introduction}
\label{Introduction}

General Relativity and the Standard Model (SM) of particle
physics are two of the most successful theories of modern
physics. Their interplay may, however, reveal phenomena
that are not apparent when gravity and particle physics
are considered separately. A particularly interesting
possibility is the existence of black holes (BHs)
surrounded by non-trivial configurations of the fundamental
fields of the SM. The present work explores this
possibility for the electroweak sector, focusing on
a remarkable non-perturbative configuration: the sphaleron.

On the one hand, BHs are a central prediction of General Relativity
\cite{Berti:2015itd}. Their astrophysical relevance is
supported by gravitational-wave observations of compact
binary mergers
\cite{LIGOScientific:2016aoc,LIGOScientific:2016sjg}
and by the horizon-scale images obtained by the
Event Horizon Telescope
\cite{EventHorizonTelescope:2019dse,EventHorizonTelescope:2022wkp}.
According to the Kerr paradigm, the exterior geometry of
an isolated, stationary astrophysical BH is expected to
be well described by the Kerr solution
\cite{Kerr:1963ud}, characterized by its mass and
angular momentum. Kerr BHs thus provide a particularly
well-motivated gravitational background in which to
investigate the behaviour of fundamental matter fields.

On the other hand, the SM provides a
remarkably successful description of the known
non-gravitational interactions. Its experimental
foundations were further strengthened by the discovery
of the Higgs boson in 2012
\cite{ATLAS:2012yve,CMS:2012qbp},
which completed the particle content of the theory.
The bosonic electroweak sector of the SM possesses
a particularly interesting classical solution:
the sphaleron
\cite{Manton:1983nd,Klinkhamer:1984di}.
This is a finite-energy, globally regular,
non-perturbative configuration of the gauge and
Higgs fields. Although unstable, the sphaleron
has important physical implications: it represents
a saddle point of the energy functional separating
topologically distinct electroweak vacua and
provides the energy barrier associated with
anomalous baryon- and lepton-number violating
transitions.

Unlike many solitonic configurations considered
in the context of BH physics, the electroweak
sphaleron is not a solution of a hypothetical
matter model. It arises within a sector of an
experimentally established fundamental theory,
whose parameters, including the Higgs boson mass,
are now measured. This makes the study of
electroweak sphalerons in BH backgrounds
particularly well motivated, even considering the BHs relevant
for significant interactions with electroweak
sphalerons have horizon sizes comparable to
the electroweak length scale and are therefore
necessarily microscopic. Their possible existence
remains hypothetical, but their study provides
a theoretically controlled setting in which
to investigate the interaction between gravity
and the non-perturbative electroweak sector
of the SM.

The simplest sphaleron is static and spherically
symmetric, the latter property being exact only
in the limit of vanishing weak mixing angle,
$\theta_{\rm W}=0$.
For a non-vanishing mixing angle, the sphaleron
is axially symmetric
\cite{Kleihaus:1991ks}.
Static axially symmetric multi-sphalerons are
also known, as well as configurations possessing
only discrete symmetries
\cite{Kleihaus:2003tn}.

Moreover, sphalerons need not be static.
As shown in~\cite{Kleihaus:2008cv,Radu:2008ta},
they can carry angular momentum, accompanied
by a non-vanishing electric charge and
magnetic dipole moment.
In particular, the angular momentum is
proportional to the electric charge, while
the net magnetic charge vanishes.
These spinning configurations provide a natural
starting point for investigating the interaction
between electroweak sphalerons and rotating BHs.

Recent precision calculations of the static sphaleron have also been reported
in~\cite{Matchev:2025ivr}, within the approximation $\theta_{\rm W}=0$.
The present work considers measured values for both $\beta$ and $\theta_{\rm W}$ and extends
the physical-parameter treatment to the axially symmetric setting, including spinning
and multi-sphaleron configurations.
With this background, the first part of the present work revisits
static and spinning (multi-)sphalerons in
flat spacetime, using the measured electroweak
parameters, including the physical Higgs boson
mass. We provide updated numerical results
for their properties, together with useful
fitting relations. These configurations will
also serve as reference solutions for the
construction of sphaleron clouds around BHs.
 
The discovery of hairy black holes in pure Einstein--Yang--Mills theory by Volkov and Gal'tsov \cite{Volkov:1989fi}
initiated a long line of work on non-Abelian black holes.
A variety of generalizations were subsequently constructed,
including BHs in spontaneously broken gauge theories, such as gravitating
monopoles and Yang--Mills--Higgs sphalerons; see the review
\jk{\cite{Volkov:1998cc}} and references therein.
An important example, particularly close to the electroweak setting considered
here, is the construction of Greene, Mathur and O'Neill
\cite{Greene:1992fw}, who obtained spherically symmetric BHs with non-trivial
Yang--Mills--Higgs fields for an $SU(2)$ theory with a Higgs doublet.
More recently, magnetically charged BHs with electroweak hair have been
obtained in the gravity-coupled Weinberg--Salam theory
\cite{Gervalle:2024yxj,Gervalle:2025awa}.
These solutions demonstrate that the electroweak vacuum can support non-trivial
field configurations in the presence of a BH horizon.

The configurations investigated here belong
to a different sector of the electroweak
theory. Rather than focusing on magnetically
charged BHs, we consider nonlinear clouds
connected to the ordinary electroweak
sphaleron and its spinning generalizations.
Our main objective is to establish whether
such configurations can coexist with a BH
horizon, how their properties depend on
the BH parameters, and, in particular,
what role is played by rotation.

For the physical values of the electroweak
couplings, the sphaleron mass is of order
$10\,{\rm TeV}$, far below the Planck scale.
Consequently, its gravitational backreaction
is negligible in the regime where the
energy stored in the matter fields remains
sufficiently small. This motivates our
use of the {\it probe limit}, in which the
electroweak field equations are solved on
a fixed BH geometry. The relevant backgrounds
are Schwarzschild for the non-rotating
case and Kerr for rotating BHs.
This approximation isolates the effect
of the BH geometry on the electroweak
fields while retaining the full nonlinear
matter dynamics.

An additional motivation comes from BH
superradiance
\cite{Brito:2015oca}.
In rotating spacetimes, suitable bosonic
field modes can extract rotational energy
from a BH. When a confining mechanism is
present, superradiant amplification may
lead to an instability, whose threshold
is associated with stationary linear clouds.
For several bosonic matter models, such
clouds are connected to families of
nonlinear configurations and hairy BHs
\cite{Herdeiro:2014goa,Herdeiro:2016tmi}.

The corresponding picture for the bosonic
electroweak theory remains comparatively
unexplored. In particular, the existence
and properties of stationary sphaleron
configurations around rotating BHs raise
several questions. Can nonlinear electroweak
sphalerons coexist with a Kerr horizon?
What is the domain of existence of the
corresponding configurations? How are
their mass, electric charge and angular
momentum affected by the BH rotation?
And can the nonlinear clouds approach
a linear configuration, as happens for
the familiar clouds associated with
superradiant thresholds?

These questions are especially interesting
because the electroweak theory contains
both massive gauge fields and a Higgs
field, with nonlinear interactions fixed
by the SM. The resulting configurations
need not follow the pattern familiar
from simpler scalar or vector models.
Related investigations in the
Proca-Higgs model can be found in~\cite{Herdeiro:2023lze}.

In this work, we construct nonlinear
electroweak sphaleron clouds around
Schwarzschild and Kerr BHs and explore
their domain of existence and physical
properties. We find families of solutions
with a non-trivial branch structure,
including configurations connected
to the corresponding flat-space sphalerons.
A notable result is that, within the
families constructed here, the nonlinear
clouds retain a finite mass gap and
do not approach a linear-cloud limit.
Thus, these solutions exhibit a
qualitatively different behaviour
from the familiar linear clouds at
superradiant thresholds.
This result concerns the sphaleron
branches studied in the present work
and does not, by itself, exclude other
electroweak configurations that might
admit a linear limit.

Our results provide a first step
towards understanding the interplay
between Kerr rotation and nonlinear
sphaleron configurations in the
bosonic sector of the SM. They also
raise further questions concerning
the possible role of backreaction,
the existence of other branches
and the stability of the resulting
configurations.

This paper is organized as follows.
In Section~\ref{section_2} we introduce the
theoretical framework, including the
axially symmetric ansatz, the quantities
of interest, the Kerr metric in
quasi-isotropic coordinates and the
boundary conditions.
Section~\ref{Static} discusses static
sphalerons and multi-sphalerons,
first in flat spacetime and subsequently
in a Schwarzschild BH background.
In Section~\ref{Rotating} we examine
spinning sphalerons in flat spacetime
and construct the corresponding
nonlinear clouds around Kerr BHs.
Section~\ref{Conclusions} summarizes
our results and discusses possible
directions for future work.
Appendix~\ref{expressions} presents
the field-strength components and
covariant derivatives associated
with the ansatz.
Finally, Appendix~\ref{ex-sol}
contains a new analytical solution
of the leading-order perturbative
Yang-Mills-Higgs problem in a
Schwarzschild BH background.

\medskip

\noindent{\bf Conventions.}
Throughout this paper, Greek indices
$\alpha,\beta,\ldots$ label spacetime coordinates,
running from $1$ to $4$, with $x^4=t$.
Early Latin indices $a,b,\ldots$ label internal
isospace components. We adopt the Einstein summation
convention and, for notational simplicity, do not
distinguish between covariant and contravariant
internal indices.

We use the metric signature $(+++-)$.
In spherical coordinates, the Minkowski line element reads
\begin{equation}
ds^2=dr^2+r^2\left(d\theta^2+
\sin^2\theta\,d\varphi^2\right)-dt^2,
\end{equation}
where $t$ denotes the time coordinate and
\begin{equation}
0\leq r<\infty,\qquad
0\leq\theta\leq\pi,\qquad
0\leq\varphi<2\pi.
\end{equation}

Unless stated otherwise, we work in Planck units,
$c=G=\hbar=1$.

\section{Theoretical setting}
\label{section_2}

We consider the bosonic sector of the SM
electroweak theory, minimally coupled to gravity.
Our main objective is to construct stationary,
nonlinear configurations of the electroweak fields
on fixed Schwarzschild and Kerr backgrounds.

We begin by introducing the field equations,
the axially symmetric ansatz and the relevant
physical quantities. We then discuss the
probe approximation, the Kerr geometry,
and the boundary conditions employed in
the numerical construction of the solutions.

\subsection{The bosonic sector of the
Weinberg-Salam theory}

The bosonic electroweak theory is based on the
gauge group $SU(2)_{\rm L}\times U(1)_{\rm Y}$.
Its field content consists of an $SU(2)$ gauge
potential
\[
V_\mu=\frac{1}{2}V_\mu^a\tau^a,
\]
where $\tau^a$ are the Pauli matrices, a
$U(1)$ gauge potential $A_\mu$, and a complex
scalar doublet $\Phi$, describing the Higgs field.

The corresponding Lagrangian density is
\begin{equation}
\label{lag}
\begin{aligned}
{\cal L}={}
-\frac{1}{2}{\rm Tr}
\left(F_{\mu\nu}F^{\mu\nu}\right)
-\frac{1}{4}f_{\mu\nu}f^{\mu\nu}
-(D_\mu\Phi)^\dagger(D^\mu\Phi)
-\lambda\left(
\Phi^\dagger\Phi-\frac{v^2}{2}
\right)^2,
\end{aligned}
\end{equation}
where
\begin{equation}
\begin{aligned}
F_{\mu\nu}
 =& {\partial_\mu V_\nu-\partial_\nu V_\mu}
 + ig[V_\mu,V_\nu],
\\
f_{\mu\nu}
&=\partial_\mu A_\nu-\partial_\nu A_\mu,
\end{aligned}
\end{equation}
are the non-Abelian and Abelian field
strengths, respectively.  
The covariant
derivative of the Higgs doublet is
\begin{equation}
D_\mu\Phi=
\left(
\partial_\mu
+ igV_\mu
+\frac{i}{2}g'A_\mu
\right)\Phi.
\end{equation}

Here, $g$ and $g'$ are the $SU(2)$ and
$U(1)$ gauge couplings, respectively,
$\lambda$ is the Higgs self-coupling,
and $v$ denotes the Higgs vacuum
expectation value.

The field equations following from
(\ref{lag}) are
\begin{eqnarray}
\label{eqsSM}
{\cal D}_\mu F^{a\nu \mu}
&=&
 \frac{ig}{2}
\left[
\Phi^\dagger\tau^aD^\nu\Phi
-(D^\nu\Phi)^\dagger\tau^a\Phi
\right],
\nonumber\\
\nabla_\mu f^{\nu \mu}
&=&
 \frac{ig'}{2}
\left[
\Phi^\dagger D^\nu\Phi
-(D^\nu\Phi)^\dagger\Phi
\right],
\\
D_\mu D^\mu\Phi
&=&
2\lambda
\left(
\Phi^\dagger\Phi-\frac{v^2}{2}
\right)\Phi.
\nonumber
\end{eqnarray}
In the limit $g'=0$, the Abelian gauge
field decouples and may consistently
be set to zero.

The stress-energy tensor is
\begin{eqnarray}
T_{\mu\nu}
&=&
2{\rm Tr}\left[
F_{\mu\alpha}F_{\nu\beta}g^{\alpha\beta}
-\frac{1}{4}g_{\mu\nu}
F_{\alpha\beta}F^{\alpha\beta}
\right]
+
f_{\mu\alpha}f_{\nu\beta}g^{\alpha\beta}
-\frac{1}{4}g_{\mu\nu}
f_{\alpha\beta}f^{\alpha\beta}
\nonumber\\
&&+
(D_\mu\Phi)^\dagger(D_\nu\Phi)
+(D_\nu\Phi)^\dagger(D_\mu\Phi)
-
g_{\mu\nu}(D_\alpha\Phi)^\dagger
(D^\alpha\Phi)
-
g_{\mu\nu}\lambda
\left(
\Phi^\dagger\Phi-\frac{v^2}{2}
\right)^2.
\label{Tik}
\label{tmunu}
\end{eqnarray}

Electroweak symmetry breaking is
characterized by the Higgs vacuum
expectation value
\begin{equation}
\label{Higgs}
\langle\Phi\rangle=
\frac{v}{\sqrt{2}}
\begin{pmatrix}
0\\1
\end{pmatrix},
\end{equation}
which gives rise to the gauge boson
and Higgs masses
\begin{equation}
\begin{aligned}
M_W&=\frac{gv}{2},
\\
M_Z&=\frac{v}{2}\sqrt{g^2+g'^2},
\\
M_H&=v\sqrt{2\lambda}.
\end{aligned}
\end{equation}

The weak mixing angle is defined by
\[
\tan\theta_{\rm W}=\frac{g'}{g},
\]
while the electromagnetic coupling is
\[
e=g\sin\theta_{\rm W}
=g'\cos\theta_{\rm W}.
\]

For the configurations considered below,
the electromagnetic potential is defined
asymptotically by
\begin{equation}
\label{emfield}
{\cal A}_\mu=
\sin\theta_{\rm W}V_\mu^3
+\cos\theta_{\rm W}A_\mu,
\end{equation}
in a gauge where the Higgs field
approaches (\ref{Higgs}) at infinity.

The corresponding electromagnetic field
strength is
\begin{equation}
{\cal F}_{\mu\nu}
=
\partial_\mu{\cal A}_\nu
-\partial_\nu{\cal A}_\mu.
\end{equation}

\subsection{The axially symmetric
Yang-Mills--Higgs--$U(1)$ ansatz}
\label{Ansatz}

To construct sphaleron configurations on
Minkowski, Schwarzschild and Kerr backgrounds,
we employ an axially symmetric ansatz for
the electroweak fields.
The general axially symmetric ansatz
introduced in~\cite{Brihaye:1994ib}
contains twenty functions: twelve
non-Abelian gauge potentials, four Abelian
potentials and four real Higgs components,
subject to residual gauge freedom.
As in previous studies of electroweak
sphalerons, we consider a consistent
ten-function truncation of this general
ansatz, incorporating additional discrete
symmetries. This choice contains the
ordinary sphaleron, its spinning
generalizations and the corresponding
axially symmetric multi-sphalerons.

For Minkowski and Schwarzschild backgrounds,
we use the parametrization
\begin{eqnarray}
\label{YM-ansatz}
V_\mu dx^\mu
&=&
\left(
\frac{\tau_r^{(n)}}{2g}H_5
+\frac{\tau_\theta^{(n)}}{2g}H_6
\right)dt
+
\left(
\frac{H_1}{r}dr+(1-H_2)d\theta
\right)
\frac{\tau_\varphi^{(n)}}{2g}
\nonumber\\
&&-
n\sin\theta
\left(
H_3\frac{\tau_r^{(n)}}{2g}
+(1-H_4)\frac{\tau_\theta^{(n)}}{2g}
\right)d\varphi,
\end{eqnarray}
for the $SU(2)$ field,
\begin{equation}
\label{M-ansatz}
A_\mu dx^\mu
=
\frac{1}{g'}
\left(
a_\varphi\sin\theta\,d\varphi
+a_0dt
\right),
\end{equation}
for the Abelian field, and
\begin{equation}
\label{H-ansatz}
\Phi=
\frac{iv}{\sqrt{2}}
\left(
\Phi_1\tau_r^{(n)}
+\Phi_2\tau_\theta^{(n)}
\right)
\begin{pmatrix}
0\\1
\end{pmatrix},
\end{equation}
for the Higgs doublet.

The matrices $\tau_r^{(n)}$,
$\tau_\theta^{(n)}$ and
$\tau_\varphi^{(n)}$ are defined
in terms of the Pauli matrices
$\vec{\tau}=(\tau_x,\tau_y,\tau_z)$
as
\begin{eqnarray}
\tau_r^{(n)}
&=&
\vec{\tau}\cdot
(\sin\theta\cos n\varphi,
\sin\theta\sin n\varphi,
\cos\theta),
\nonumber\\
\tau_\theta^{(n)}
&=&
\vec{\tau}\cdot
(\cos\theta\cos n\varphi,
\cos\theta\sin n\varphi,
-\sin\theta),
\nonumber\\
\tau_\varphi^{(n)}
&=&
\vec{\tau}\cdot
(-\sin n\varphi,
\cos n\varphi,0).
\end{eqnarray}

The positive integer $n$ is the winding
number, with $n=1$ corresponding to
the ordinary sphaleron and $n>1$
to multi-sphaleron configurations.

All ten functions
\[
H_i,\quad a_\varphi,\quad a_0,
\quad\Phi_1,\quad\Phi_2
\]
depend only on $(r,\theta)$.
Consequently, the stress-energy tensor
is stationary and axially symmetric.

We further impose reflection symmetry
with respect to the equatorial plane,
$\theta\rightarrow\pi-\theta$.
The functions
\[
H_2,H_4,H_6,\Phi_1,a_\varphi,a_0
\]
are even under this reflection,
whereas
\[
H_1,H_3,H_5,\Phi_2
\]
are odd.

\medskip

The ansatz is axially symmetric in
the gauge-covariant sense: a rotation
around the symmetry axis can be
compensated by an internal gauge
transformation
\cite{Forgacs:1979zs,Brihaye:1994ib}.
For the $SU(2)$ potential,
\begin{equation}
\label{gauge_rotation}
{\cal L}_\varphi V_\mu
={\cal D}_\mu\Psi,
\end{equation}
where ${\cal L}_\varphi$ denotes
the Lie derivative and
\[
{\cal D}_\mu X
=
\partial_\mu X+ig[V_\mu,X]
\]
is the gauge-covariant derivative
in the adjoint representation.

For the ansatz (\ref{YM-ansatz}),
\begin{equation}
\label{psi_lie}
\Psi=
\frac{n}{2g}\tau_z
=
\frac{n}{2g}
\left(
\cos\theta\,\tau_r^{(n)}
-\sin\theta\,\tau_\theta^{(n)}
\right).
\end{equation}

It follows that
\begin{equation}
\label{relation1}
F_{\mu\varphi}
=
{\cal D}_\mu{\cal W},
\qquad
{\cal W}=V_\varphi-\Psi.
\end{equation}

The Higgs doublet and Abelian potential
satisfy
\begin{equation}
\label{relation2}
{\cal L}_\varphi\Phi
=
-\frac{in}{2}\Phi
-ig\Psi\Phi,
\qquad
{\cal L}_\varphi A_\mu=0.
\end{equation}

These relations will be useful in
characterizing the angular momentum
and imposing regularity at the Kerr
horizon. Explicit expressions for
the field strengths and Higgs
covariant derivatives are given
in Appendix~\ref{expressions}.

\medskip

The ansatz retains a residual
Abelian gauge freedom,
\begin{equation}
\label{gauge}
U=
\exp\left(
\frac{i}{2}
\tau_\varphi^{(n)}
\Gamma(r,\theta)
\right),
\end{equation}
where $\Gamma(r,\theta)$ is real.

Under this transformation,
\begin{equation}
\label{gt1}
\begin{aligned}
H_1&\longrightarrow
H_1-r\partial_r\Gamma,
\\
H_2&\longrightarrow
H_2+\partial_\theta\Gamma.
\end{aligned}
\end{equation}

Thus, $H_1$ and $H_2$ transform
as the components of an effective
two-dimensional gauge field,
while $(H_3+\cot\theta,H_4)$
forms a scalar doublet.

Following Ref.~\cite{Kleihaus:1997mn},
we fix this residual freedom by
imposing
\begin{equation}
\label{gc1}
r\partial_rH_1-\partial_\theta H_2=0.
\end{equation}
This condition is incorporated
into the numerical scheme.

For Kerr backgrounds we retain
the Higgs ansatz (\ref{H-ansatz}),
while adopting a slightly modified
parametrization of the gauge
potentials, given in
Eqs.~(\ref{YM-ansatzK}) and
(\ref{M-ansatzK}) below.
The gauge-covariant symmetry
relations and residual gauge
condition remain unchanged.

\subsection{Physical quantities and scalings}
\label{quant}

The sphaleron configurations are characterized
by their mass-energy, angular momentum,
electric charge and magnetic dipole moment.

In the probe approximation, the mass-energy
and angular momentum carried by the matter
fields outside the horizon are obtained from
the corresponding components of the
stress-energy tensor,
\begin{eqnarray}
\label{M}
M
&=&
-\int d^3x\sqrt{-g}\,T_t^t
=
-4\pi\int_{r_H}^{\infty}dr
\int_0^{\pi/2}d\theta\,
\sqrt{-g}\,T_t^t,
\\
\label{J}
J
&=&
\int d^3x\sqrt{-g}\,T_\varphi^t
=
4\pi\int_{r_H}^{\infty}dr
\int_0^{\pi/2}d\theta\,
\sqrt{-g}\,T_\varphi^t.
\end{eqnarray}

Here, $r_H=0$ for globally regular
configurations in Minkowski spacetime.
The second equalities follow from
axial symmetry and reflection symmetry
across the equatorial plane.

The rotating configurations also
carry an electric charge and
a magnetic dipole moment, which
can be extracted from the
asymptotic electromagnetic fields.

For the solutions considered here,
the large-$r$ behaviour of the
electric gauge potentials is
\begin{eqnarray}
H_5
&=&
\left(V-\frac{Q}{r}\right)
\cos\theta+\cdots,
\\
H_6
&=&
\left(V-\frac{Q}{r}\right)
\sin\theta+\cdots,
\\
a_0
&=&
V-\frac{Q}{r}+\cdots,
\label{asympa0}
\\
a_\varphi
&=&
\frac{\zeta}{r}\sin\theta
+\cdots.
\label{asymp}
\end{eqnarray}

The constant $V$ specifies
the electrostatic potential
at infinity, while $Q$ and
$\zeta$ are determined by
the numerical solution.

The total electric charge is
obtained from the electromagnetic
flux at spatial infinity,
\begin{equation}
\label{Qel}
{\cal Q}
=
\frac{1}{4\pi}
\oint_{S^2}
{}^\star{\cal F}_{\theta\varphi}
\,d\theta\,d\varphi
=
\frac{Q}{e}.
\end{equation}

Similarly, the magnetic dipole
moment $\mu$ is defined by
\begin{equation}
{\cal A}_\varphi
\longrightarrow
\frac{\mu\sin^2\theta}{4\pi r},
\end{equation}
which yields
\begin{equation}
\mu=\frac{4\pi\zeta}{e}.
\end{equation}

\medskip

The natural length scale of the
electroweak sphaleron is
\[
L_S=\frac{1}{gv}
=\frac{1}{2M_W}.
\]
We therefore introduce the
dimensionless radial coordinate
\begin{equation}
\label{scaling}
\hat r=
\frac{r}{L_S}
=gv\,r.
\end{equation}
The same scaling is applied
to the Kerr mass parameter
$M_K$ and the horizon
radius $r_H$.

The dimensionless Higgs
field is defined by
\begin{equation}
\label{scalingH}
\hat\Phi=\frac{\Phi}{v},
\end{equation}
while the electric potentials
are rescaled according to
\begin{equation}
\label{scalingHi}
\hat H_5=\frac{H_5}{gv},
\qquad
\hat H_6=\frac{H_6}{gv},
\qquad
\hat a_0=\frac{a_0}{gv}.
\end{equation}
In particular,
\[
\hat V=\frac{V}{gv}.
\]

{In terms of the dimensionless variables, we define the numerical asymptotic coefficients by}
\begin{eqnarray}
\hat H_5
&=&
{\left(\hat V-\frac{Q^{({\rm num})}}{\hat r}\right)\cos\theta+\cdots,}
\\
\hat H_6
&=&
{\left(\hat V-\frac{Q^{({\rm num})}}{\hat r}\right)\sin\theta+\cdots,}
\\
\hat a_0
&=&
{\hat V-\frac{Q^{({\rm num})}}{\hat r}+\cdots,}
\\
a_\varphi
&=&
{\frac{\zeta^{({\rm num})}}{\hat r}\sin\theta+\cdots.}
\label{asympnum}
\end{eqnarray}
Comparison with Eqs.~(\ref{asympa0}) and (\ref{asymp}) gives $Q^{({\rm num})}=Q$ and $\zeta^{({\rm num})}=gv\,\zeta$.

After these rescalings,
the matter equations depend
on the two dimensionless
electroweak parameters
\begin{eqnarray}
\label{beta}
\beta^2
&=&
\frac{\lambda}{g^2}
=
\frac{1}{8}
\left(\frac{M_H}{M_W}\right)^2
\simeq0.3029,
\\
\label{theta}
\tan\theta_{\rm W}
&=&
\frac{g'}{g}
\simeq0.5488.
\end{eqnarray}

For rotating configurations,
the electrostatic potential
$\hat V$ provides an additional
continuous parameter.
The existence of finite-energy,
localized solutions restricts
its range to
\[
-\frac{1}{2}\leq
\hat V\leq\frac{1}{2}.
\]

Unless stated otherwise,
all numerical results are
obtained for the physical
values of $\beta$ and
$\theta_{\rm W}$.
The only exception is the
spherical approximation
considered in
Section~\ref{approx}.

The physical mass, angular momentum, electric charge and magnetic dipole moment are recovered from the dimensionless numerical quantities through 
\begin{equation}
M=\frac{4\pi v}{g}M^{(\mathrm{num})},
\qquad
J=\frac{4\pi}{g^2}J^{(\mathrm{num})},
\qquad
\mathcal{Q}=\frac{1}{e}Q^{(\mathrm{num})},
\qquad
\mu=\frac{4\pi}{e g v}\zeta^{(\mathrm{num})}.
\label{MJQmu}
\end{equation}

In what follows, we display
dimensionless quantities
unless explicitly stated
otherwise, suppressing
hats and the superscript
$({\rm num})$ when no
confusion can arise.

\subsection{The probe approximation
and gravitational backreaction}
\label{backreaction}

Our main objective is to investigate how
a BH horizon and its rotation affect
nonlinear electroweak sphaleron
configurations.

For physical electroweak parameters,
the gravitational coupling of the
sphaleron is extremely small.
Introducing the dimensionless
coupling
\begin{equation}
\alpha^2=Gv^2,
\end{equation}
one obtains
\begin{equation}
\alpha\simeq2\times10^{-17}.
\end{equation}

Consequently, the deformation of
the spacetime geometry produced
by an ordinary electroweak
sphaleron is negligible.

This motivates the probe
approximation adopted throughout
the present work: the electroweak
field equations are solved on
a prescribed solution of the
vacuum Einstein equations,
without including the
backreaction of the matter
fields on the geometry.

For static BHs, the background
is Schwarzschild, while for
rotating BHs we employ Kerr.

The probe approximation allows
us to isolate the influence
of the BH geometry on the
electroweak fields while
retaining their full nonlinear
interactions.

It should be stressed, however,
that the smallness of $\alpha$
does not guarantee the validity
of the approximation for
arbitrarily large matter
configurations. In particular,
the approximation may cease
to be self-consistent along
branches on which the
energy stored in the fields
becomes sufficiently large.
We return to this issue
when discussing the
excited branches.

\subsection{The Kerr black hole background}
\label{KerrBH}

The rotating BH background is described
by the Kerr solution of the vacuum
Einstein equations.
For the numerical construction of
the electroweak configurations,
we employ quasi-isotropic coordinates,
rather than the standard
Boyer-Lindquist parametrization.

This coordinate system is particularly
convenient for the boundary-value
problem, since regularity at the
event horizon can be implemented
through Dirichlet or Neumann
conditions on the matter functions.

We write the metric as
\begin{equation}
\label{metric}
\begin{aligned}
ds^2={}
-f\,dt^2
+\frac{m}{f}
\left(
dr^2+r^2d\theta^2
\right)
+
\frac{l}{f}r^2\sin^2\theta
\left(
d\varphi-\frac{\omega}{r}dt
\right)^2,
\end{aligned}
\end{equation}
where $f,m,l$ and $\omega$
depend on $(r,\theta)$.

For Kerr, the metric functions
take the form
\begin{eqnarray}
\label{Kerr}
f
&=&
\left(
1-\frac{r_H^2}{r^2}
\right)^2
\frac{F_1}{F_2},
\nonumber \qquad
l
=
\left(
1-\frac{r_H^2}{r^2}
\right)^2,
\nonumber\\
m
&=&
\left(
1-\frac{r_H^2}{r^2}
\right)^2
\frac{F_1^2}{F_2}, \qquad 
\omega
=
\frac{
2M_K\sqrt{M_K^2-4r_H^2}
}{r^2}
\frac{
1+\frac{M_K}{r}
+\frac{r_H^2}{r^2}
}{F_2},
\end{eqnarray}
where
\begin{eqnarray}
\label{functions-Kerr1}
F_1
&=&
\frac{2M_K^2}{r^2}
+
\left(
1-\frac{r_H^2}{r^2}
\right)^2
+
\frac{2M_K}{r}
\left(
1+\frac{r_H^2}{r^2}
\right)
-
\frac{M_K^2-4r_H^2}{r^2}
\sin^2\theta,
\\
\label{functions-Kerr2}
F_2
&=&
\left[
\frac{2M_K^2}{r^2}
+
\left(
1-\frac{r_H^2}{r^2}
\right)^2
+
\frac{2M_K}{r}
\left(
1+\frac{r_H^2}{r^2}
\right)
\right]^2
-
\left(
1-\frac{r_H^2}{r^2}
\right)^2
\frac{M_K^2-4r_H^2}{r^2}
\sin^2\theta.
\end{eqnarray}

The solution is characterized
by the Kerr mass parameter
$M_K$ and the isotropic
horizon radius $r_H$,
satisfying
\[
M_K\geq2r_H.
\]

The angular momentum,
horizon area and Hawking
temperature are
\begin{eqnarray}
J_K
&=&
M_K\sqrt{M_K^2-4r_H^2},
\nonumber\\
A_H
&=&
8\pi M_K(M_K+2r_H),
\nonumber\\
T_H
&=&
\frac{r_H}{
2\pi M_K(M_K+2r_H)
}.
\end{eqnarray}

The horizon is generated
by the Killing vector
\begin{equation}
\label{chi}
\chi=
\partial_t+
\Omega_H\partial_\varphi,
\end{equation}
where the horizon angular
velocity is
\begin{equation}
\label{rel1}
\Omega_H=
\frac{
\sqrt{M_K^2-4r_H^2}
}{
2M_K(M_K+2r_H)
}.
\end{equation}

For the numerical study it
is convenient to characterize
the Kerr background by
$(r_H,\Omega_H)$.

Introducing the dimensionless
variables
\[
x=\frac{r_H}{M_K},
\qquad
U=\Omega_Hr_H,
\]
Eq.~(\ref{rel1}) yields
\begin{equation}
x^3-\frac{1}{2}x^2
+4U^2x+2U^2=0.
\end{equation}

For a given $U$ in the
allowed range, this equation
possesses two physical roots,
corresponding to two branches
of Kerr backgrounds.

Defining
\[
C(U)=
\frac{
1-288U^2
}{
(1-48U^2)^{3/2}
},
\]
these roots can be written as
\begin{eqnarray}
\label{roots}
x_{\rm up}
&=&
\frac{1}{6}
\left[
1+
2\sqrt{1-48U^2}
\cos\left(
\frac{\arccos C(U)}{3}
\right)
\right],
\nonumber\\
x_{\rm low}
&=&
\frac{1}{6}
\left[
1+
2\sqrt{1-48U^2}
\cos\left(
\frac{\arccos C(U)+4\pi}{3}
\right)
\right].
\end{eqnarray}
The two branches merge
at the maximum value
\begin{equation}
\label{Umax}
U_{\rm max}
=
\frac{1}{
(3+\sqrt{5})
\sqrt{2(1+\sqrt{5})}
}
\simeq0.0750708.
\end{equation}

At this point, the dimensionless
Kerr spin is
\[
j_K=\frac{J_K}{M_K^2}
\simeq0.786151.
\]

The branch $x_{\rm up}$
extends from the Schwarzschild
limit to this critical
configuration, whereas
$x_{\rm low}$ connects
the critical configuration
to the extremal Kerr limit. 
{So the upper-$x$ branch has lower dimensionless spin $j_K$,
and vice versa.} 

\medskip

For Kerr backgrounds,
we retain the Higgs ansatz
(\ref{H-ansatz}) and use
the following parametrization
of the gauge potentials:
\begin{eqnarray}
\label{YM-ansatzK}
V_\mu dx^\mu
&=&
\left(
\frac{\tau_r^{(n)}}{2g}H_5
+
\frac{\tau_\theta^{(n)}}{2g}H_6
\right)dt
+
\left(
\frac{H_1}{r}dr
+(1-H_2)d\theta
\right)
\frac{\tau_\varphi^{(n)}}{2g}
\nonumber\\
&&-
n\sin\theta
\left(
H_3\frac{\tau_r^{(n)}}{2g}
+
(1-H_4)
\frac{\tau_\theta^{(n)}}{2g}
\right)
\times
\left(
d\varphi-\frac{\omega}{r}dt
\right),
\\
\label{M-ansatzK}
A_\mu dx^\mu
&=&
\frac{1}{g'}
\left[
a_\varphi\sin\theta
\left(
d\varphi-\frac{\omega}{r}dt
\right)
+a_0dt
\right].
\end{eqnarray}

This parametrization is adapted
to the rotating geometry
and simplifies the regularity
conditions at the horizon.
It reduces to the previous
ansatz in the absence of
background rotation.

\subsection{Boundary conditions
and numerical construction}
\label{numerics}

\subsubsection{Boundary conditions}
\label{BCS}

The electroweak field equations,
together with the axially symmetric
ansatz, lead to a coupled system
of nonlinear partial differential
equations.

The boundary conditions are chosen
to ensure regularity at the origin
or event horizon, regularity on
the symmetry axis, and finite
energy and angular momentum.

For static configurations,
the electric potentials vanish,
\[
H_5=H_6=a_0=0,
\]
and the system reduces to seven
coupled equations.

In Minkowski spacetime,
the boundary conditions are
\begin{eqnarray}
\label{BC-static-flat}
H_1=H_3=a_\varphi
=\Phi_1=\Phi_2
&=&0,
\nonumber\\
H_2=H_4&=&1,
\qquad r=0,
\nonumber\\[2mm]
H_1=H_3=a_\varphi
=\Phi_2&=&0,
\nonumber\\
H_2=H_4&=&-1,
\qquad
\Phi_1=1,
\qquad r\rightarrow\infty,
\nonumber\\[2mm]
H_1=H_3=a_\varphi
=\Phi_2&=&0,
\nonumber\\
\partial_\theta H_2
=\partial_\theta H_4
=\partial_\theta\Phi_1
&=&0,
\qquad\theta=0,\pi.
\end{eqnarray}

For Schwarzschild backgrounds,
the conditions at infinity and
on the symmetry axis remain
unchanged.

At the event horizon,
the regularity conditions are
\begin{equation}
\label{BCh-S}
\begin{aligned}
\partial_rH_2
=\partial_rH_3
=\partial_rH_4&=0,
\\
\partial_r\Phi_1
=\partial_r\Phi_2&=0,
\qquad r=r_H.
\end{aligned}
\end{equation}

For spinning configurations,
all ten functions in the
ansatz are retained.

The magnetic and Higgs
functions obey the same
boundary conditions as
in the static case.

In Minkowski spacetime,
the electric potentials
satisfy
\begin{eqnarray}
\label{BC-rotating-flat}
\partial_r a_0&=&0,
\nonumber\\
\sin\theta\,H_5
+\cos\theta\,H_6&=&0,
\nonumber\\
\cos\theta\,\partial_rH_5
-\sin\theta\,\partial_rH_6
&=&0,
\qquad r=0,
\nonumber\\[2mm]
a_0&=&V,
\nonumber\\
H_5&=&V\cos\theta,
\nonumber\\
H_6&=&V\sin\theta,
\qquad r\rightarrow\infty,
\nonumber\\[2mm]
H_6
=\partial_\theta H_5
=\partial_\theta a_0
&=&0,
\qquad\theta=0,\pi.
\end{eqnarray}

For Kerr backgrounds,
the asymptotic and axial
conditions remain unchanged.
At the horizon, the magnetic
and Higgs functions satisfy
the conditions of the
static BH problem,
while the electric potentials
obey
\begin{equation}
\label{KErr-BC}
\begin{aligned}
H_5&=
n\Omega_H\cos\theta,
\\
H_6&=
-n\Omega_H\sin\theta,
\\
a_0&=\Omega_H,
\qquad r=r_H.
\end{aligned}
\end{equation}

These conditions ensure
that the gauge potentials
are compatible with the
horizon generator
(\ref{chi}) and that
the corresponding
electrostatic potentials
are constant on the horizon
\cite{Kleihaus:2002ee,Kleihaus:2016rgf}.

Regularity on the symmetry
axis additionally requires
\[
H_2=H_4,
\qquad\theta=0,\pi,
\]
a condition verified by
the numerical solutions.

The reflection symmetries
specified above allow the
angular domain to be
restricted to
$0\leq\theta\leq\pi/2$.

Finally, the asymptotic
Higgs configuration is
\begin{equation}
\label{HiggsA}
\Phi\longrightarrow
\frac{iv}{\sqrt{2}}
\begin{pmatrix}
\sin\theta\,e^{-in\varphi}\\
-\cos\theta
\end{pmatrix}.
\end{equation}

It can be transformed
to the standard
electroweak vacuum
(\ref{Higgs}) by
\[
U=
\exp\left(
-\frac{i\pi}{2}
\tau_r^{(n)}
\right).
\]

Under this transformation,
the gauge functions transform
according to
\[
(H_1,H_2,H_3,H_4,H_5,H_6)
\longrightarrow
(-H_1,-H_2,H_3,-H_4,H_5,-H_6).
\]

\subsubsection{Numerical method}
\label{num1}

The axially symmetric solutions
are obtained by solving the
coupled nonlinear elliptic
equations using a finite-difference
Newton-Raphson scheme
\cite{SCHONAUER1989279,
SCHONAUER1990279,
SCHONAUER2001473}.
The radial coordinate is
compactified according to
\begin{equation}
x=
\frac{r-r_H}{c+r},
\end{equation}
where $c$ is a positive
numerical parameter,
usually chosen as $c=1$.
This maps the exterior
domain
$r_H\leq r<\infty$
to $0\leq x<1$,
allowing the asymptotic
boundary conditions
to be imposed at a
finite coordinate value.

The initial configuration
is the spherically symmetric
static sphaleron, obtained
independently by a shooting
method.
The axially symmetric
solutions are then generated
through numerical continuation,
varying the weak mixing angle,
winding number, electrostatic
potential or BH parameters
in sufficiently small steps.

For the solutions presented
here, the typical estimated
relative numerical error
in the matter functions is
of order $10^{-4}$.
As an additional consistency
check, we evaluate a virial
identity derived following
the approach of
Ref.~\cite{Gervalle:2024yxj},
finding deviations of
comparable magnitude.

\section{Static solutions}
 \label{Static}
We begin with the non-rotating sector, which provides both the reference
configurations and the starting points for the construction of sphaleron
clouds around rotating BHs. We first revisit globally regular
sphalerons in Minkowski spacetime, using the measured electroweak parameters,
and then introduce a Schwarzschild horizon. Our emphasis is on the
continuation between flat-space solutions and BH clouds, and on the
additional branches that arise in the presence of a horizon.

The ordinary sphaleron and its finite-mixing-angle generalization are
well established \cite{Manton:1983nd,Klinkhamer:1984di,Kleihaus:1991ks,Kunz:1992uh},
as are axially symmetric multi-sphalerons
\cite{Kleihaus:1994yj,Kleihaus:1994tr}. We restrict attention to this
sphaleron sector; sphaleron--antisphaleron pairs, chains and vortex rings,
which require an additional integer in the gauge-field ansatz, will not
be considered \cite{Kleihaus:2008gn}.

\subsection{Static (multi-)sphalerons on Minkowski spacetime}

\subsubsection{The spherical approximation: $\theta_{\rm W}=0$}
\label{approx}

At vanishing weak mixing angle, $g'=0$, the Abelian field consistently
decouples. For a purely magnetic configuration with winding number $n=1$,
the sphaleron can then be described by the spherically symmetric $SU(2)$
Yang--Mills--Higgs ansatz
\begin{equation}
\label{spherical-ansatz}
\begin{aligned}
H_1=H_3=H_5=H_6=a_\varphi=a_0&=0,\\
H_2=H_4&=w(r),\\
\Phi_1&=\phi(r),\qquad \Phi_2=0.
\end{aligned}
\end{equation}
Here and below, $r$ denotes the dimensionless radius introduced in
Section~\ref{quant}. The field equations reduce to two coupled ordinary
differential equations,
\begin{equation}
\label{eqKM1}
w''=\frac{w(w^2-1)}{r^2}+\frac14(1+w)\phi^2,
\end{equation}
\begin{equation}
\label{eqKM2}
\frac{d}{dr}\left(r^2\phi'\right)
=\frac12\phi(1+w)^2
+\beta^2r^2\phi(\phi^2-1).
\end{equation} 

Regularity at the origin and approach to the electroweak vacuum at infinity
require
\begin{equation}
w(0)=1,\qquad \phi(0)=0,\qquad
w(\infty)=-1,\qquad \phi(\infty)=1.
\end{equation}
The corresponding dimensionless energy density is
\begin{equation}
\label{T44-KM}
T_{tt}=\frac{w'^2}{r^2}
+\frac{(1-w^2)^2}{2r^4}
+\frac12\phi'^2
+\frac{\phi^2(1+w)^2}{4r^2}
+\frac{\beta^2}{4}(\phi^2-1)^2.
\end{equation}

\begin{figure}[h]
    \centering
    \includegraphics[width=0.65\linewidth]{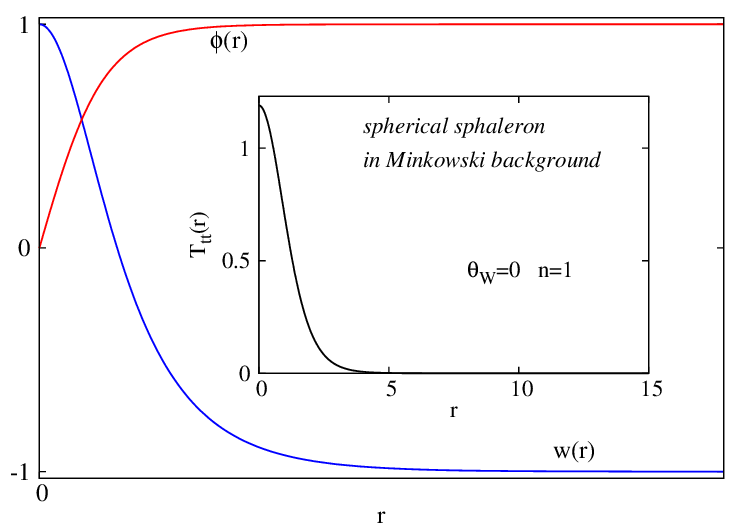}
    \caption{The profile of the spherically symmetric  sphaleron  with
    ${\rm \theta}_W=0$ and the physical value of
    the $\beta$-parameter (\ref{beta})
    is shown as a function of the radial coordinate.}
    \label{sphaleron-flat}
\end{figure}

Figure~\ref{sphaleron-flat} shows the spherical solution at the physical
Higgs-sector parameter $\beta$ of Eq.~(\ref{beta}). The gauge and Higgs
profiles are monotonic, with the energy concentrated on the electroweak
length scale. We find the dimensionless mass $M=1.91612$. The corresponding
finite-mixing-angle result below is $M=1.89873$, a difference of about
$0.9\%$; the spherical approximation is therefore useful for the mass,
but does not capture the physical electromagnetic dipole field. 

For comparison, the simple profiles
\begin{equation}
w(r)=\frac{2}{\cosh(ar)}-1,\qquad
\phi(r)=\tanh(br),\qquad
a=0.577,\quad b=0.528,
\end{equation}
proposed in~\cite{Tye:2015tva}
{as an approximate solution of eqs. (\ref{eqKM1}), (\ref{eqKM2}),}
yield $M=1.91682$, close to the
numerical value.

\subsubsection{Finite weak mixing angle and physical electroweak parameters}

At nonzero weak mixing angle, the ordinary sphaleron sources an Abelian
magnetic field and is no longer exactly spherical. It retains axial and
equatorial-reflection symmetry, so that the gauge and Higgs functions
depend on both $r$ and $\theta$. We solve the full axially symmetric
boundary-value problem with the physical values of $\theta_{\rm W}$ and
$\beta$ specified in Section~\ref{quant}.

\begin{figure}[h]
    \centering
    \begin{minipage}{0.49\textwidth}
        \centering
        \includegraphics[width=\linewidth]{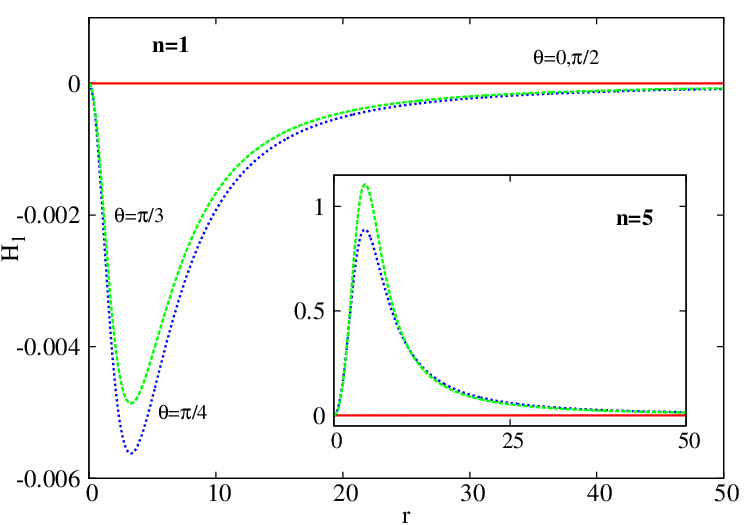}
    \end{minipage}
    \hfill
    \begin{minipage}{0.49\textwidth}
        \centering
        \includegraphics[width=\linewidth]{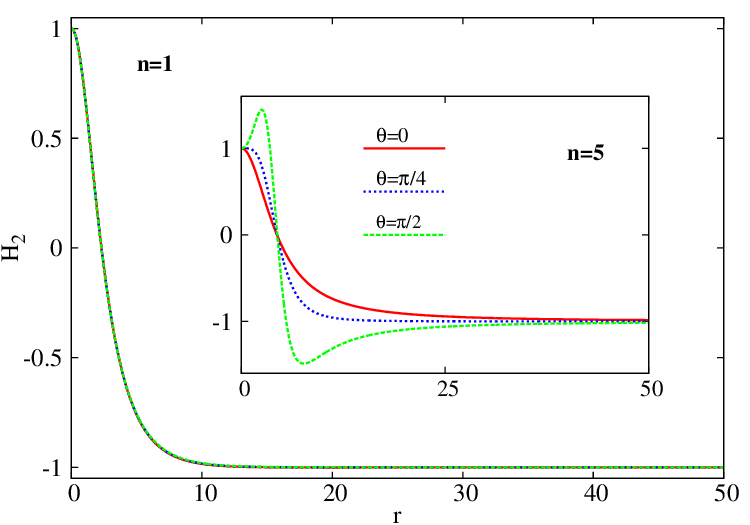}
    \end{minipage}

    \par\vspace{0.5cm} 

    \begin{minipage}{0.49\textwidth}
        \centering
        \includegraphics[width=\linewidth]{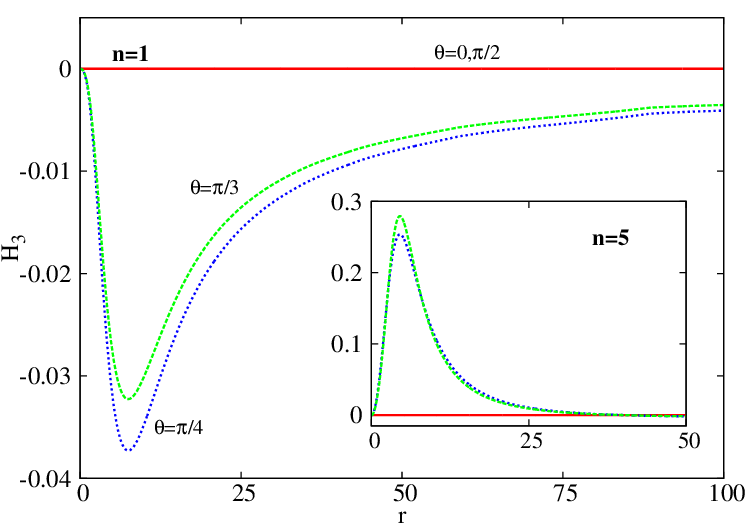}
    \end{minipage}
    \hfill
    \begin{minipage}{0.49\textwidth}
        \centering
        \includegraphics[width=\linewidth]{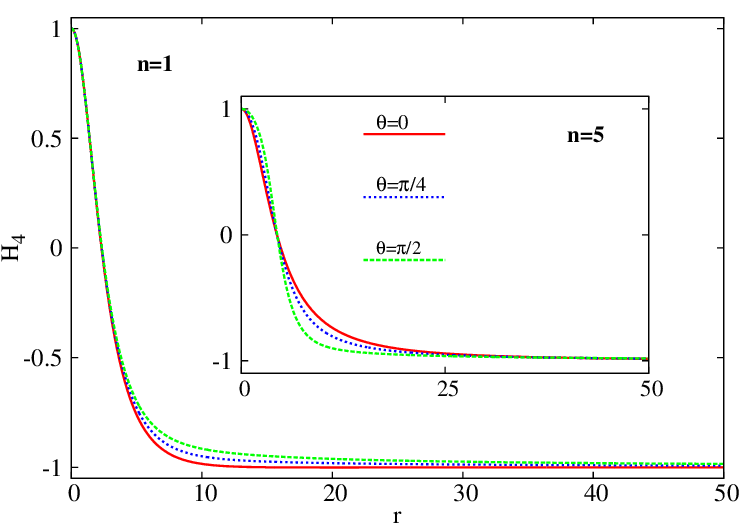}
    \end{minipage}
    \caption{The profiles of the Yang-Mills magnetic potentials $H_i$ 
    which enter the sphaleron Ansatz
    are shown as a function of the radial coordinate for several different angles. 
    The insets show the profiles 
    of the multi-sphaleron solution with winding number $n=5$.
    }
    \label{Hi-static}
\end{figure}

\begin{figure}[h]
    \centering
    \begin{minipage}{0.49\textwidth}
        \centering
        \includegraphics[width=\linewidth]{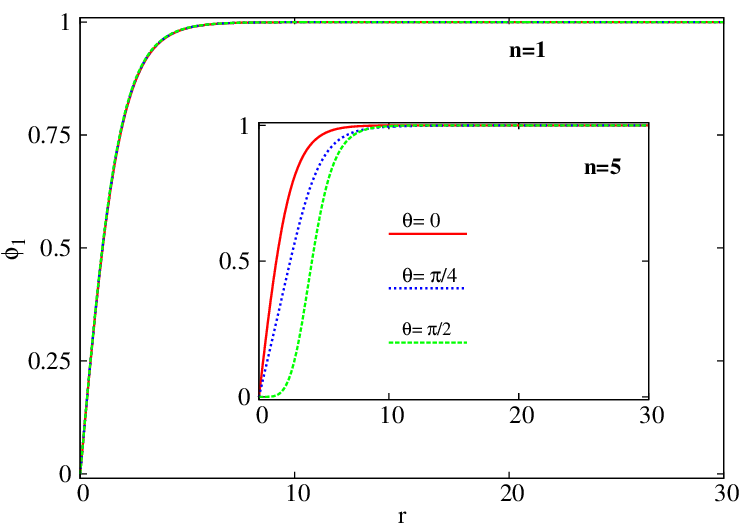}
    \end{minipage}
    \hfill
    \begin{minipage}{0.49\textwidth}
        \centering
        \includegraphics[width=\linewidth]{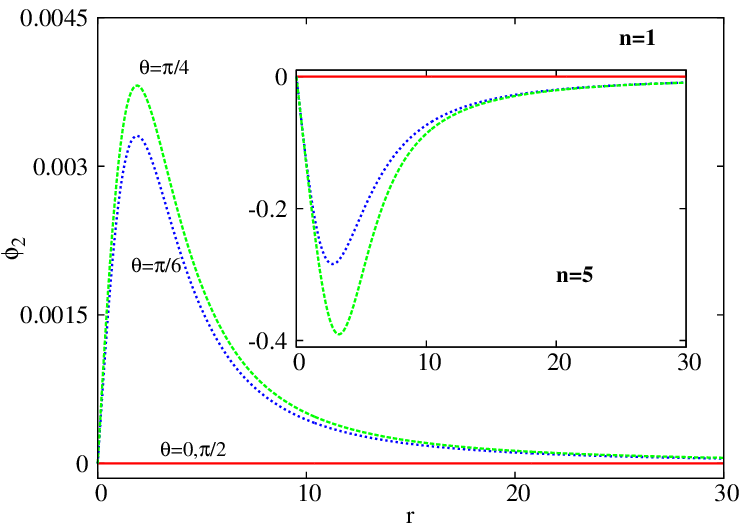}
    \end{minipage}

    \par\vspace{0.5cm} 

    \begin{minipage}{0.49\textwidth}
        \centering
        \includegraphics[width=\linewidth]{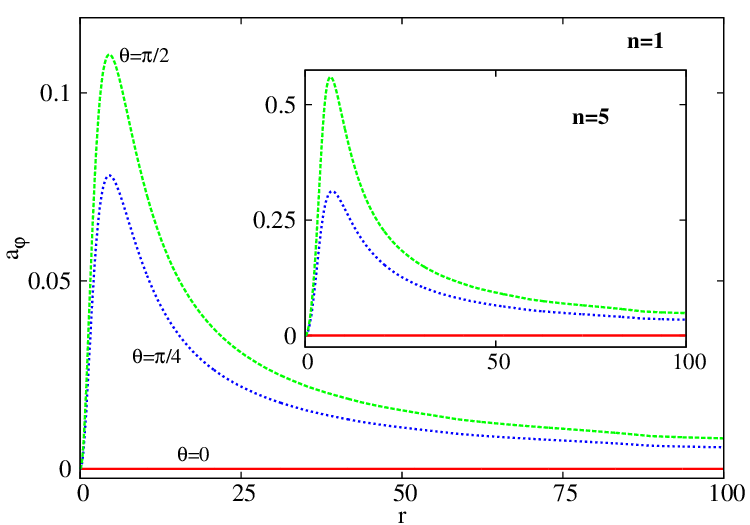}
    \end{minipage}
    \hfill
    \begin{minipage}{0.49\textwidth}
        \centering
        \includegraphics[width=\linewidth]{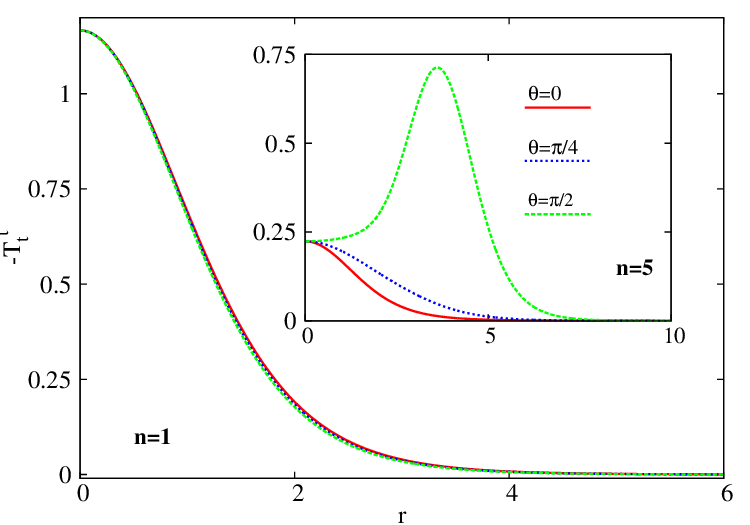}
    \end{minipage}
    \caption{Same as Figure \ref{Hi-static}
    for the Higgs functions $\Phi_{1,2}$, the U(1) magnetic potential $a_\varphi$
   and the energy density represented by $-T_t^t$. 
    }
    \label{Zi-static}
\end{figure}

\begin{figure}[h]
\vspace{-0.5cm}
\centering
\mbox{ 
\hspace*{-1.5cm}\includegraphics[height=.4\textheight, angle =0]{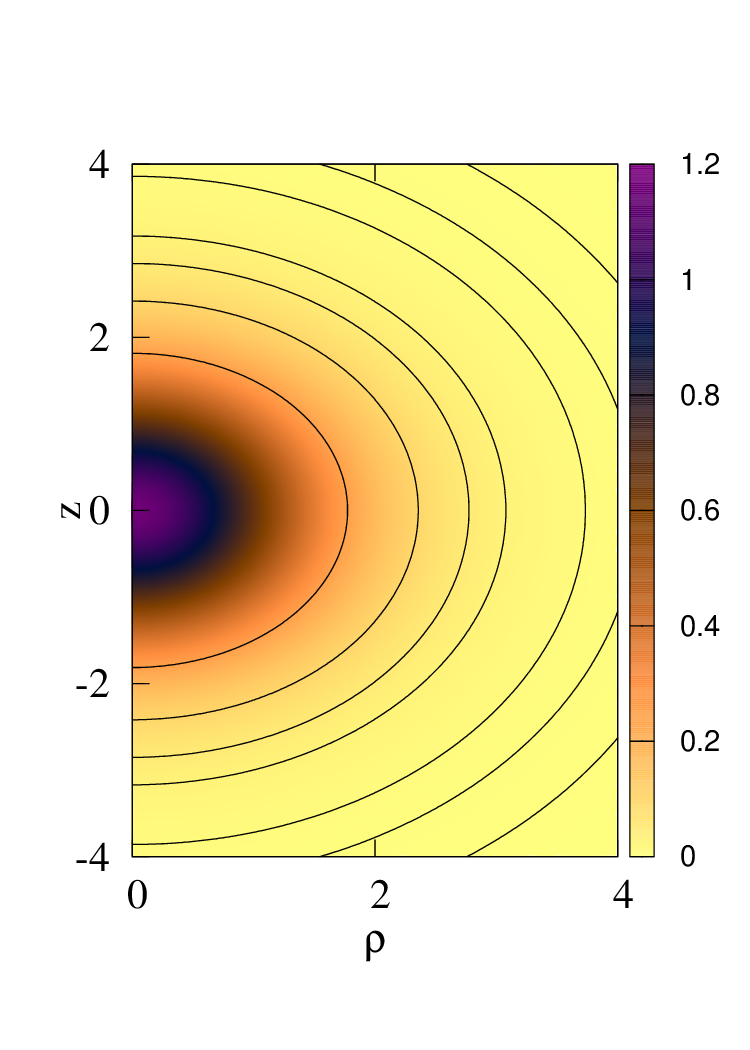}
\hspace*{-0.5cm}\includegraphics[height=.4\textheight, angle =0]{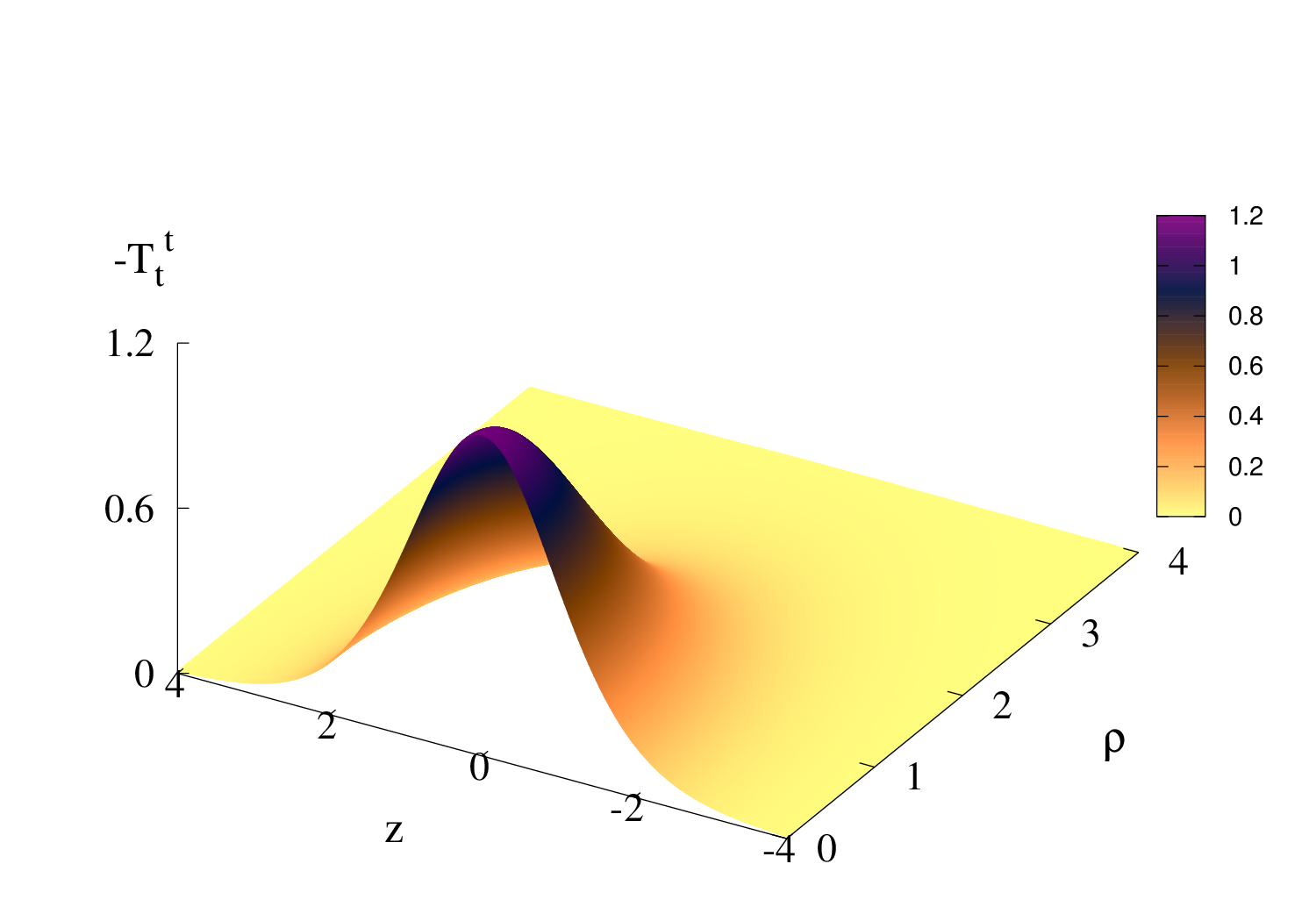}
}
\vspace{-0.5cm}
\caption{
 The mass-energy density of the sphaleron ($n=1$) 
 with the physical value of $(\theta_{\rm W},\beta)$ is shown as heat map (left) and as a 3D plot (right). 
}
\label{fig1}
\end{figure}

The gauge and Higgs profiles are displayed in
Figs.~\ref{Hi-static} and \ref{Zi-static}, while
Fig.~\ref{fig1} shows the energy density in cylindrical coordinates
$(\rho,z)$. At the physical mixing angle the deformation of the energy
density away from spherical symmetry is modest. Larger deformations at
increased mixing angle were previously found in~\cite{Kleihaus:1991ks,Kunz:1992uh}. Despite its small energetic
contribution, the Abelian field is physically relevant: the sphaleron has
an electromagnetic dipole moment, whose nonperturbative value is
close to the perturbative result of~\cite{Klinkhamer:1984di}.

The same physical-parameter calculation can be extended to the axially
symmetric multi-sphaleron sector by taking $n>1$. In contrast to the
$n=1$ solution, these configurations become progressively more deformed
as the winding number increases. The $n=5$ profiles are shown in the
insets of Figs.~\ref{Hi-static} and \ref{Zi-static}.

\begin{figure}[t!]
\vspace{-0.5cm}
\centering
\mbox{ 
\hspace*{-1.5cm}\includegraphics[height=.4\textheight, angle =0]{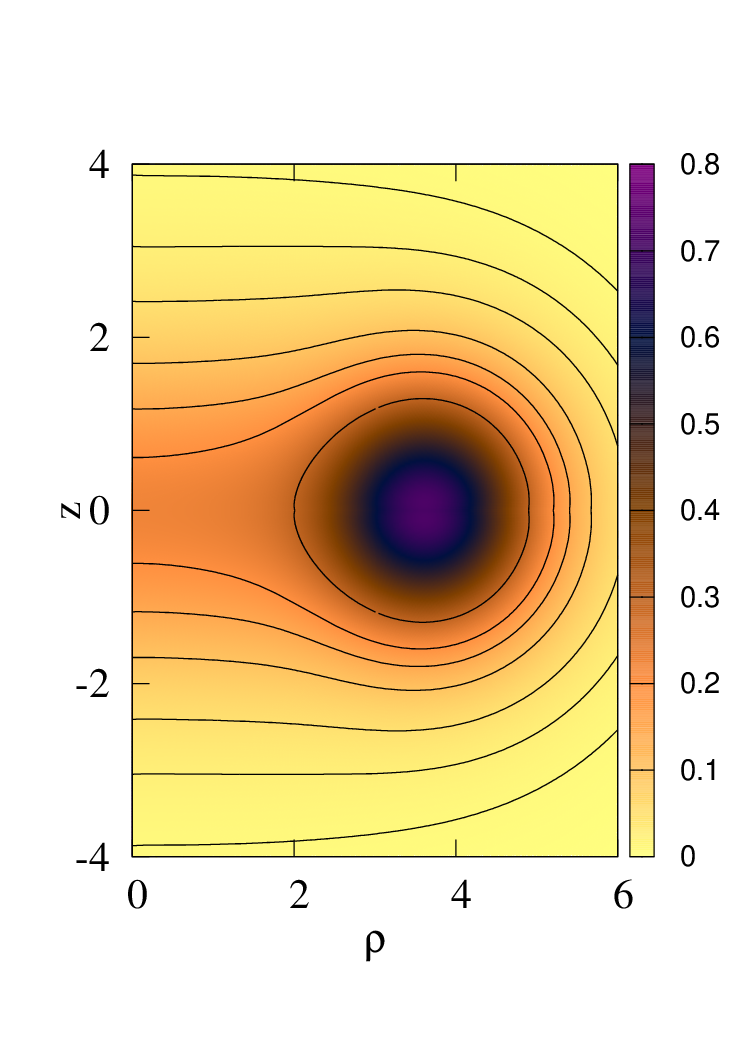}
\hspace*{-0.5cm}\includegraphics[height=.4\textheight, angle =0]{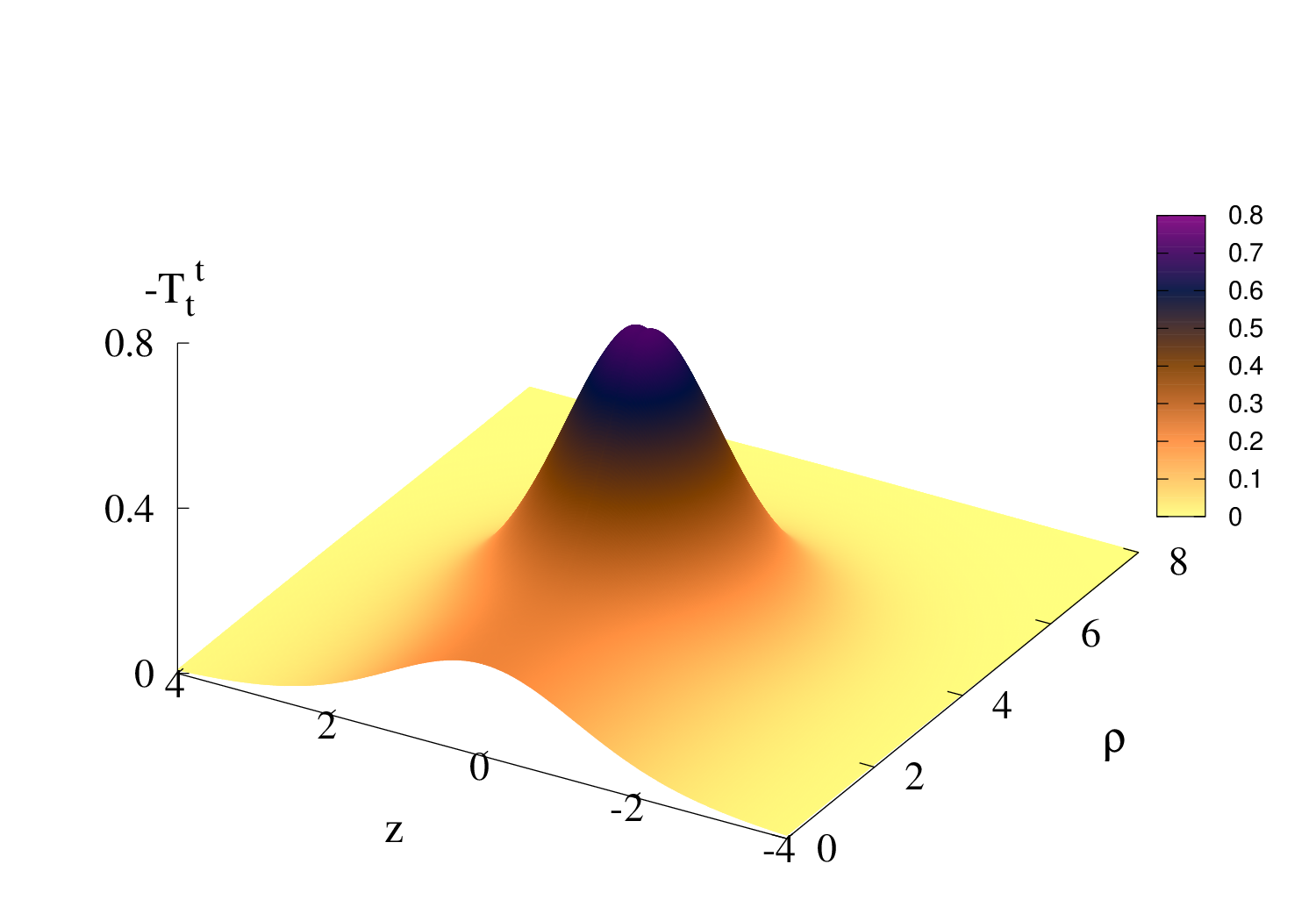}
}
\vspace{-0.5cm}
\caption{Same as Fig.~\ref{fig1} for the $n=5$ multi-sphaleron. 
}
\label{fig2}
\end{figure}

Figure~\ref{fig2} illustrates the characteristic toroidal energy
distribution: for this example the maximum lies on a circle in the
equatorial plane, rather than on the symmetry axis. The data in
Table~\ref{tab:table1} quantify the effect of the winding number on the
mass, the separate field contributions, the magnetic moment and a
characteristic size.

\begin{table}[t]
\centering
\begin{tabular}{c|ccccc}
$n$ & $M$ & $M_{\rm YM}/M$ & $M_{\rm H}/M$ & {$\zeta^{({\rm num})}$} 
& $R_{99}$\\
\hline
$1$ & $1.89873$ & $0.55256$ & $0.43926$ & $0.77413$ & $5.62082$\\
$2$ & $3.81380$ & $0.57175$ & $0.41693$ & $1.58769$ & $6.25674$\\
$3$ & $5.80508$ & $0.58222$ & $0.40397$ & $2.50378$ & $6.86318$
\end{tabular}
\caption{Dimensionless properties of static sphalerons and
multi-sphalerons at the physical values of $\theta_{\rm W}$ and $\beta$.
The symbols $M_{\rm YM}$ and $M_{\rm H}$ denote the non-Abelian gauge and
Higgs contributions to the total matter energy; $R_{99}$ encloses $99\%$
of that energy.}
\label{tab:table1}
\end{table}

The non-Abelian gauge and Higgs sectors account for nearly all the mass;
the remaining Abelian contribution is of order one percent for the
windings tabulated here. Since the fields have no sharp surface, we
characterize their extent by $R_{99}$, the radius of the sphere enclosing
$99\%$ of the total energy. It grows more slowly than the winding number
in the cases shown, remaining of order a few electroweak length scales.

Using Eq.~(\ref{MJQmu}) with $g\simeq0.65$ and $v\simeq246\,{\rm GeV}$,
the $n=1$ mass is approximately $9.03\,{\rm TeV}$, consistent with
Refs.~\cite{Tye:2015tva,Matchev:2025ivr}.
The tabulated multi-sphalerons satisfy $M(n)>nM(1)$; their energy exceeds
that of $n$ widely separated $n=1$ sphalerons. This comparison by itself
does not determine a force between separated configurations.

\begin{figure}[h]
    \centering
    \includegraphics[width=0.65\linewidth]
    {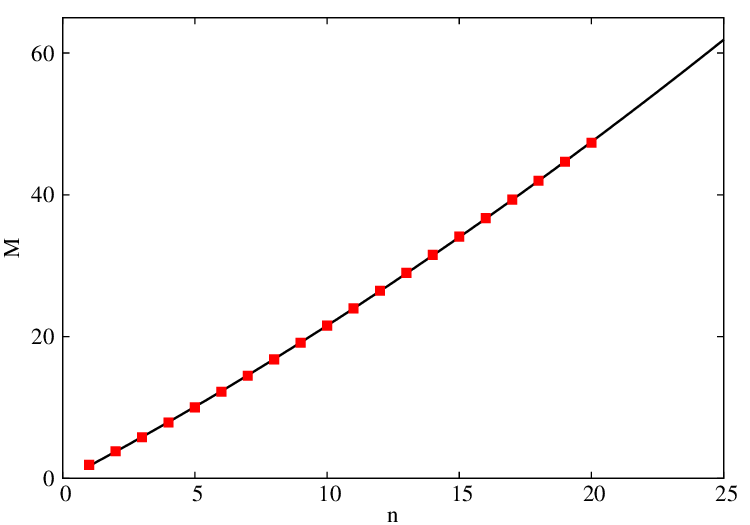}
    \caption{Multi-sphaleron mass $M$
as a function of the winding number $n$. 
Calculated data points are shown in red, while the black line indicates the fit (\ref{fit}), extrapolated to larger $n$.  
}
    \label{Ms(n)}
\end{figure}

To summarize the winding-number dependence, we computed masses for
$1\leq n\leq20$ and fitted them by the empirical relation
\begin{equation}
\label{fit}
M(n)=a+bn+cn^2\bigl(1-e^{-dn}\bigr),
\end{equation}
with
\begin{equation}
a=-0.25351,\qquad b=1.99569,\qquad
c=0.0195943,\qquad d=0.280702 
\end{equation}
(see also Figure  \ref{Ms(n)}).
This fit is a compact interpolation of the numerical results over the
stated range; its extrapolation to larger $n$ should not be interpreted
as an established large-winding asymptotic law.

\subsection{Sphaleron clouds in a Schwarzschild black hole background}
\label{Sphaleron_Schwarzschild}

We now replace the regular origin by a Schwarzschild horizon, retaining
the full nonlinear electroweak matter equations in the probe limit.
Schwarzschild is the nonrotating member of the Kerr family, obtained by
setting $J_K=0$, or equivalently $M_K=2r_H$. In isotropic coordinates,
\begin{equation}
\label{sph}
\begin{aligned}
ds^2={}-\left(\frac{1-r_H/r}{1+r_H/r}\right)^2dt^2+\left(1+\frac{r_H}{r}\right)^4
\left[dr^2+r^2\left(d\theta^2+
\sin^2\theta\,d\varphi^2\right)\right].
\end{aligned}
\end{equation}
The Schwarzschild mass and horizon area are
\begin{equation}
M_{\rm Sch}=2r_H,\qquad A_H=64\pi r_H^2.
\end{equation}
All radii and masses in this subsection are dimensionless unless stated
otherwise. The parameter $r_H$ is the \emph{isotropic coordinate radius},
not the areal radius. The latter is
\begin{equation}
R_H=\sqrt{\frac{A_H}{4\pi}}=4r_H.
\end{equation}

\begin{figure}[t!]
\vspace{-0.3cm}
\centering
\mbox{ 
\hspace*{-1.5cm}\includegraphics[height=.4\textheight, angle =0]{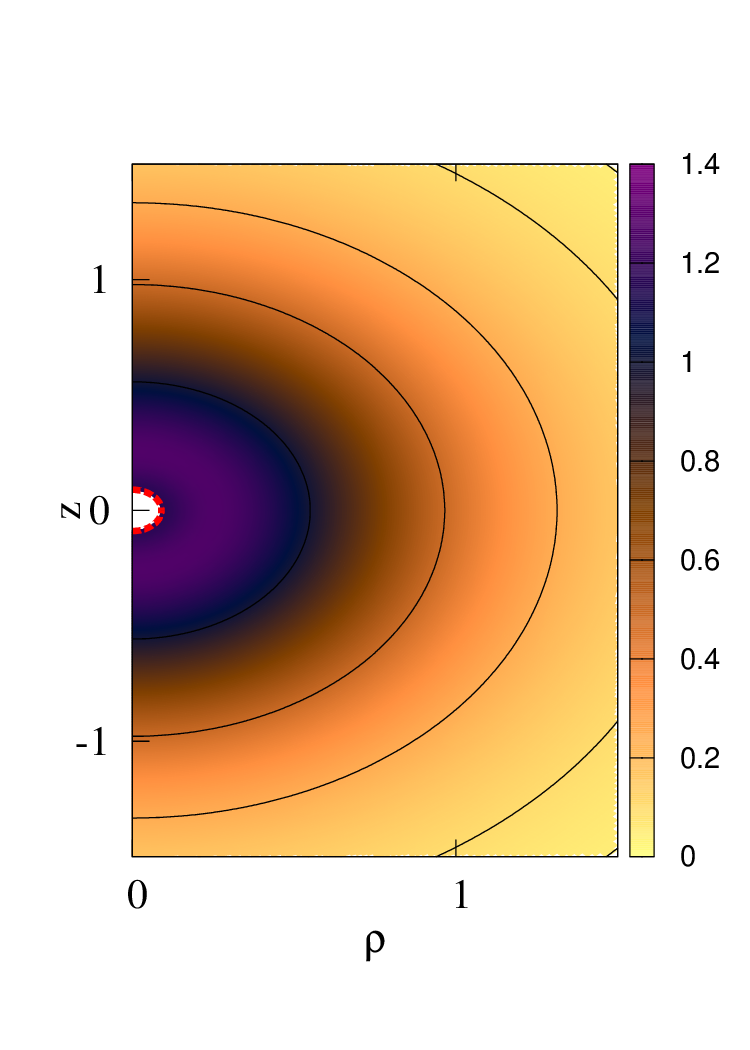}
\hspace*{ 0.5cm}\includegraphics[height=.4\textheight, angle =0]{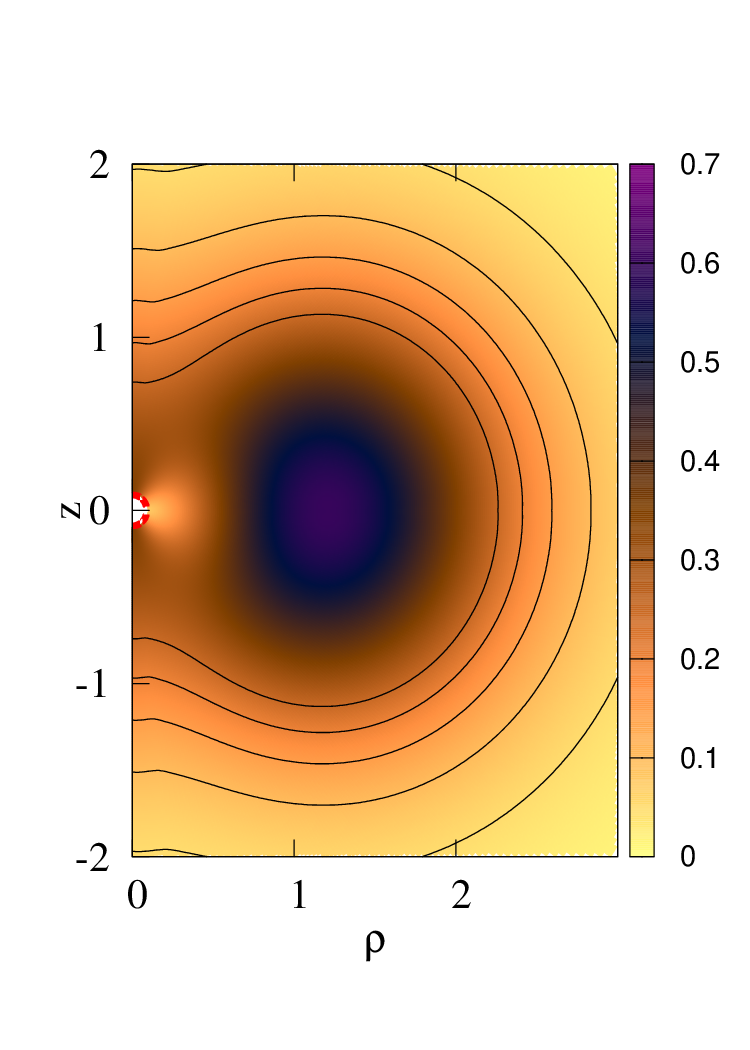}
}
\caption{ The mass-energy density of a typical fundamental branch solution with $n=1$
(left) and  $n=2$ (right) in a Schwarzschild BH
background.
The horizon ($r_H=0.09$) is indicated by the red dotted curve. 
}
\label{fig4}
\end{figure}

The distributions of matter energy for representative fundamental-branch
solutions with $n=1,2$ and $r_H=0.09$ are shown in Fig.~\ref{fig4}.
For these small horizons the profiles resemble the corresponding
flat-space sphalerons: replacing the regular centre by an event horizon
does not eliminate the surrounding nonlinear configuration.

\begin{figure}[!ht]
\hbox to\linewidth{\hss%
	\resizebox{9cm}{7cm}{\includegraphics{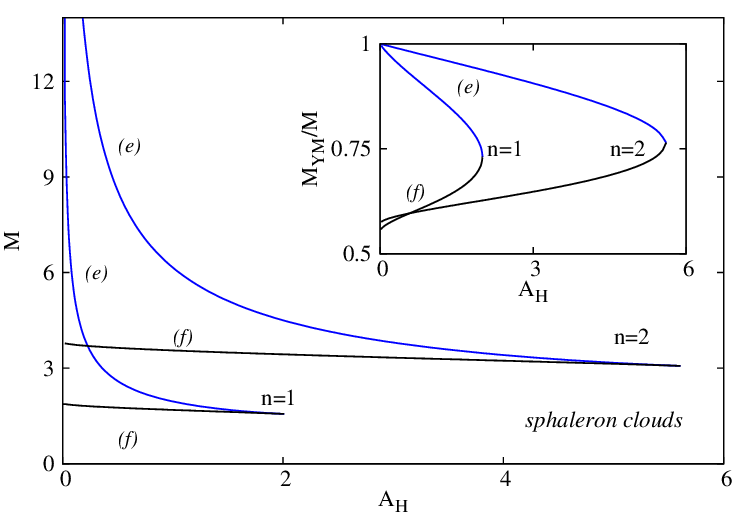}} 
\hss}
\caption{\small  
The total mass of static (multi-)sphaleron clouds in
a Schwarzschild BH background
is shown as a function of event horizon area 
$A_H= 64 \pi r_H^2$
for $n=1,2$ families of solutions.
The inset shows the ratio of the mass stored in the Yang-Mills fields.
}
\label{sph-Schw}
\end{figure}

The dependence of the \emph{matter} mass on horizon area is displayed
in Fig.~\ref{sph-Schw}. Two features are apparent. First, within the
families constructed here, the horizon size is bounded above.
For $n=1$ we find $r_H^{\rm max}\simeq0.1$, while the limiting size
increases moderately with $n$. In terms of the areal radius, the
$n=1$ endpoint corresponds to $R_H^{\rm max}\simeq0.4L_S\simeq
5\times10^{-19}\,{\rm m}$ at the physical electroweak scale.
Thus the relevant backgrounds are microscopic BHs; this scale should
not be confused with that of the observed astrophysical Kerr population.

Second, for each winding number shown, the solutions form two branches
that meet at the maximal horizon size. The \emph{fundamental} branch,
labelled $(f)$ in Fig.~\ref{sph-Schw}, is continuously connected to the
regular flat-space sphaleron as $r_H\to0$. Its matter mass decreases
modestly as the horizon grows, reaching a value about $20\%$ below its
flat-space value near the branch endpoint.

The \emph{excited} branch, labelled $(e)$, behaves differently: its
dimensionless matter mass grows without bound as $r_H\to0$. This is a
nonuniform limit of the nonlinear boundary-value problem, not a second
regular flat-space sphaleron. Numerically, the product
$r_HM^{({\rm num})}$ tends to a finite nonzero value on this branch.

The limiting behaviour can be understood from the dimensionless
combination $\hat r_H=gvr_H$. At fixed physical electroweak parameters,
shrinking the horizon drives $\hat r_H\to0$. Equivalently, the
\emph{limiting equations} can be exposed formally by taking $v\to0$ at
fixed nonzero horizon size, so that the symmetry-breaking scale becomes
negligible. The latter is a mathematical scaling argument, not a change
of the measured electroweak parameters used in the numerical results.
Within our boundary conditions, the limiting configuration has a trivial
Higgs field and Abelian sector and is described by a pure $SU(2)$
Yang--Mills field on Schwarzschild.

For $n=1$, our numerical results approach the known spherically symmetric
Yang--Mills solution of Ref.~\cite{Boutaleb-Joutei:1979hdt}, whose
isotropic-coordinate form is
\begin{equation}
\label{exact-sol}
H_1=H_3=0,\qquad H_2=H_4=w(r),
\end{equation}
with
\begin{equation}
w(r)=-\frac{\displaystyle\sqrt3+
\frac{1-\sqrt3}{4rr_H}(r+r_H)^2}
{\displaystyle3-
\frac{1-\sqrt3}{4rr_H}(r+r_H)^2}.
\end{equation}
Consistently, the numerical fraction $M_{\rm YM}/M$ approaches unity
along the excited branch, as seen in the inset of
Fig.~\ref{sph-Schw}. We have also obtained axially symmetric $n>1$
generalizations of this limiting Yang--Mills sector numerically.
Their Abelian fields likewise vanish in the limiting solutions.

The relation to the exact $n=1$ Yang--Mills background motivates a
perturbative treatment of the Higgs contribution for small $\hat r_H$.
Appendix~\ref{ex-sol} presents an analytic solution of the leading-order
Higgs problem around the configuration (\ref{exact-sol}).

Finally, the divergent excited-branch mass must be interpreted within
the probe approximation. As the matter energy increases, backreaction
will eventually become important, despite the small electroweak
gravitational coupling. Related Einstein--Yang--Mills and
Einstein--Yang--Mills--Higgs systems exhibit regular, finite-mass
horizonless limits in place of the corresponding fixed-background
divergence, including the Bartnik--McKinnon solitons
\cite{Bartnik:1988am,Greene:1992fw}. Whether the same detailed
regularization occurs for the full electroweak configurations studied
here remains a question for the backreacting theory.

\section{Rotating solutions}
\label{Rotating}

We now turn to the central problem of this work: nonlinear electroweak
sphaleron clouds on a rotating Kerr background. We first establish the
flat-space spinning configurations at the physical electroweak parameters.
These supply a reference family and clarify the relation between the
matter angular momentum and electric charge. We then introduce a Kerr
horizon, examine how the resulting families differ from their flat-space
counterparts, and map their domain of existence.

\subsection{Rotating sphalerons in Minkowski spacetime}

Spinning electroweak sphalerons were constructed in
Refs.~\cite{Radu:2008ta,Kleihaus:2008cv}; see also
Refs.~\cite{Ibadov:2010ei,Ibadov:2010hm} for further developments.
Here we revisit them using the physical values of
$\theta_{\rm W}$ and $\beta$ and extend the numerical results to
multi-sphalerons with $n=2,3$. Besides providing updated quantitative
information, these solutions furnish the horizonless limits of the
fundamental Kerr-cloud branches.

An important property of the globally regular spinning solutions is
that their total angular momentum and electric charge are proportional.
To exhibit the origin of this relation, we write the angular-momentum
density entering Eq.~(\ref{J}) as
\begin{equation}
\label{J1}
\begin{aligned}
T_\varphi^{\ t}={}
2\operatorname{Tr}\!\left(F_{\mu\varphi}F^{\mu t}\right)
+f_{\mu\varphi}f^{\mu t}+(D_\varphi\Phi)^\dagger(D^t\Phi)
 +(D^t\Phi)^\dagger(D_\varphi\Phi).
\end{aligned}
\end{equation}
Using the gauge-covariant axial symmetry relations
(\ref{relation1})--(\ref{relation2}) and the field equations
(\ref{eqsSM}), this expression becomes a total divergence,
\begin{equation}
\label{J2}
T_\varphi^{\ t}
=\frac{1}{\sqrt{-g}}\partial_\mu\left\{
\sqrt{-g}\left[
2\operatorname{Tr}\bigl({\cal W}F^{\mu t}\bigr)
+\left(A_\varphi-\frac{n}{g'}\right)f^{\mu t}
\right]\right\}.
\end{equation}
For a globally regular configuration, there is no inner boundary.
Consequently, $J$ is determined entirely by a surface integral at
spatial infinity \cite{Kleihaus:2008cv,Radu:2008ta}:
\begin{equation}
\label{Joint}
J=\lim_{r\to\infty}\int_{S^2}
\sqrt{-g}\left[
2\operatorname{Tr}\bigl({\cal W}F^{rt}\bigr)
+\left(A_\varphi-\frac{n}{g'}\right)f^{rt}
\right]d\theta\,d\varphi.
\end{equation}
The asymptotic behaviour (\ref{asympa0}) and the charge definition
(\ref{Qel}) then yield
\begin{equation}
\label{JQrel}
\frac{J}{4\pi}
=nQ\left(\frac{1}{g^2}+\frac{1}{g'^2}\right)
=\frac{nQ}{e^2}
=\frac{n{\cal Q}}{e}.
\end{equation}
Thus the angular momentum of a regular spinning sphaleron is tied to
its electromagnetic charge. The explicit normalization of $Q$ and
${\cal Q}$ should be read consistently with the conventions of
Section~\ref{quant}.

The static solution is recovered when the electrostatic parameter at
infinity vanishes. More generally, the field equations imply an
integral identity for the electric sector,
\begin{equation}
\label{Me1}
\begin{aligned}
 \operatorname{Tr}\left(F_{\mu t}F^{\mu t}\right)
 +\frac12 f_{\mu t}f^{\mu t}
 +(D_t\Phi)^\dagger(D^t\Phi)=\frac{1}{\sqrt{-g}}\partial_\mu
\left\{\sqrt{-g}\left[
 \operatorname{Tr}\bigl(V_tF^{\mu t}\bigr)
+\frac12 A_tf^{\mu t}\right]\right\}.
\end{aligned}
\end{equation}
For flat-space regular configurations, the inner contribution vanishes,
so the integrated electric term is controlled by its asymptotic value,
\begin{equation}
\label{Me1a}
M_{\rm el}=-\lim_{r\to\infty}\int_{S^2}
\sqrt{-g}\left[
 \operatorname{Tr}\bigl(V_tF^{rt}\bigr)
+\frac12 A_tf^{rt}\right]d\theta\,d\varphi.
\end{equation}
With the conventions for the asymptotic potentials used here, this was
written as
\begin{equation}
\label{Me2}
\frac{M_{\rm el}}{4\pi}=\frac12\frac{V}{e}{\cal Q}.
\end{equation}
The flat-space identity, together with the sign-definiteness of the
electric energy, excludes a nontrivial electric sector when its
asymptotic potential vanishes. This conclusion is specific to globally
regular configurations: a BH horizon supplies an additional
boundary contribution, as discussed below.

\begin{figure}[t]
\centering
\includegraphics[width=.6\linewidth]{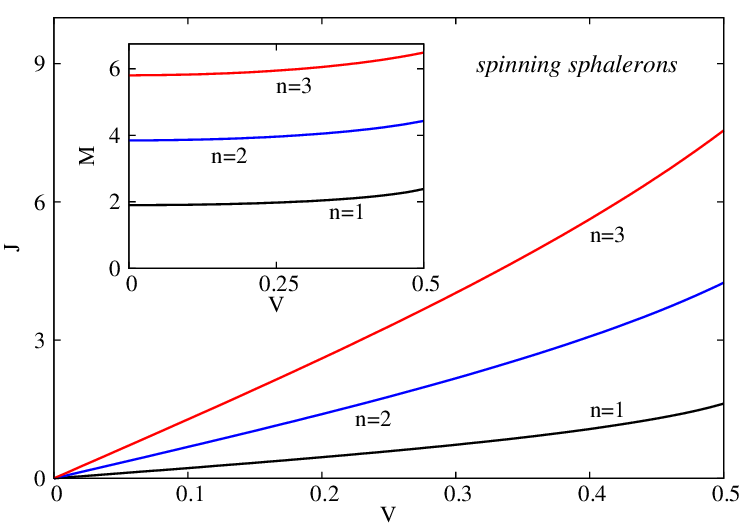}
\caption{Mass and angular momentum of flat-space spinning
(multi-)sphalerons as functions of the asymptotic electrostatic
parameter $V$, for the indicated winding numbers $n$.}
\label{spinning-sph1}
\end{figure}

The spinning branches emerge continuously from the corresponding
static sphalerons as $V$ is increased from zero, as illustrated in
Fig.~\ref{spinning-sph1}. In the dimensionless conventions of
Section~\ref{quant}, localization restricts the asymptotic parameter
to $|V|\leq M_W=1/2$. Indeed, the large-radius equations contain a
massive-field fall-off of the form
\begin{equation}
\frac{\exp\!\left(-\sqrt{\frac14-V^2}\,r\right)}{r},
\end{equation}
which ceases to be exponentially decaying when $|V|>1/2$.
Near this boundary the mass and angular momentum increase along the
branch studied, although the change in mass remains moderate.

{For $n=1$, the numerical results can be represented by numerical fits. Since the mass is invariant under $V\to-V$, whereas the angular momentum changes sign, we employ }
\begin{equation}
\label{fitM}
M(V)=M_0+\sum_{k=1}^{4} a_{2k}V^{2k},
\
J(V)=\sum_{k=0}^{3} b_{2k+1}V^{2k+1},
\end{equation}
with
\begin{align*}
M_0 = 1.89873,\\
(a_2,a_4,a_6,a_8) =
(1.15953,\ 1.72751,\ -5.47394,\ 44.5785),\\
(b_1,b_3,b_5,b_7) =
(2.18187,\ 3.14852,\ -8.90460,\ 52.2820).
\end{align*}
>%
The corresponding coefficients of determination are
$R_M^2\simeq0.999969$ and $R_J^2\simeq0.999989$.

\begin{figure}[h!]  
	\centering  
	\includegraphics[width=.38\linewidth]{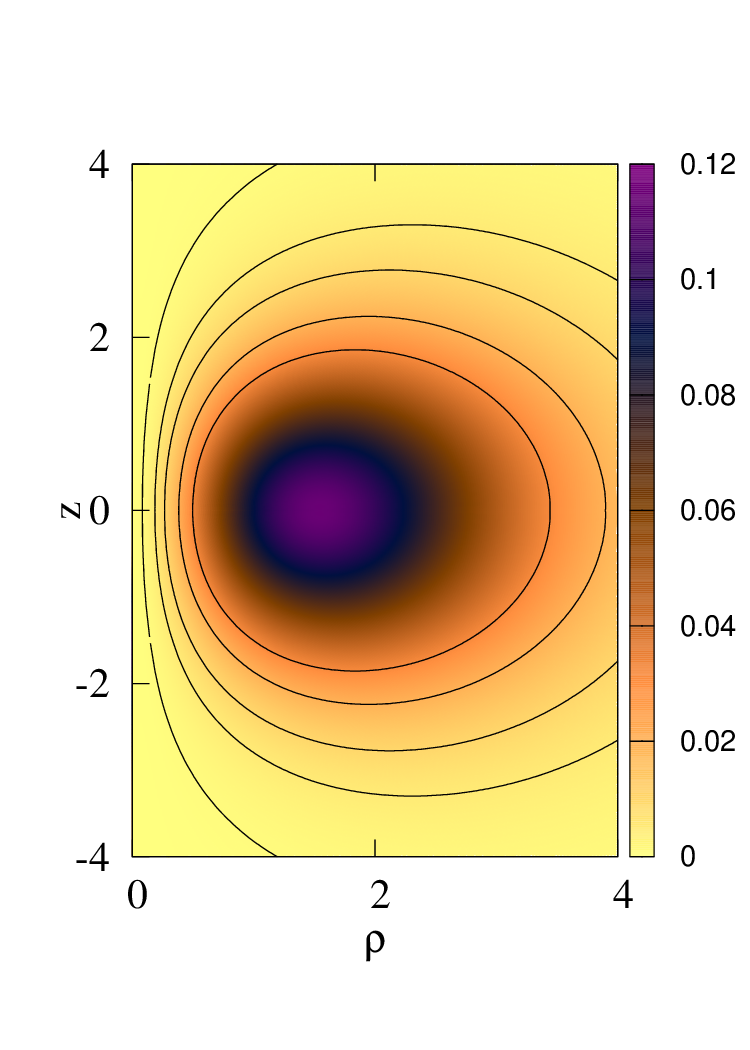}\hfill  
	\adjustbox{raise=1.5cm}{\includegraphics[width=.60\linewidth]{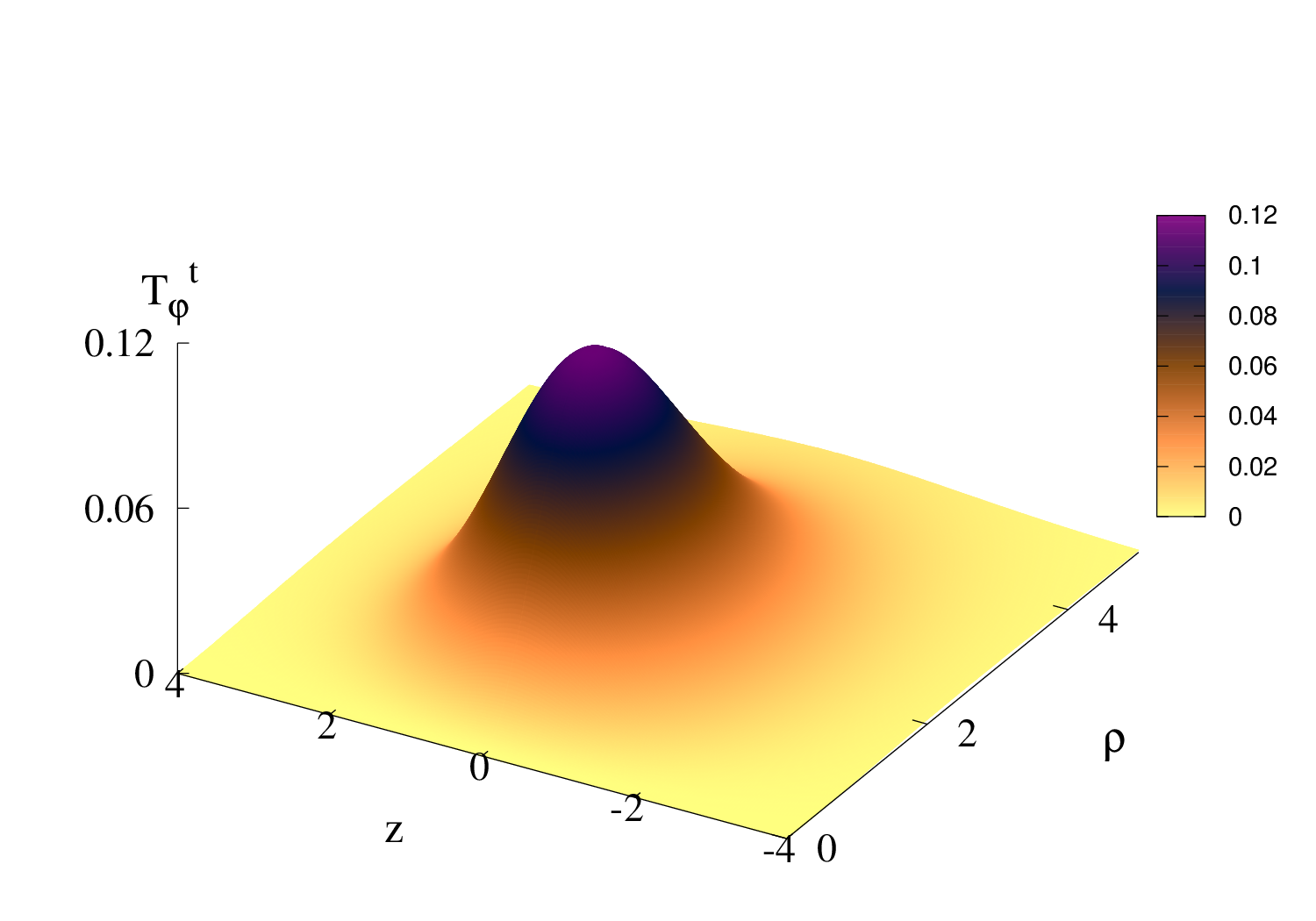}}  
	\caption{Angular-momentum density of a flat-space spinning sphaleron  
		with $n=1$, $V=0.4$ and physical $(\theta_{\rm W},\beta)$, shown as a  
		heat map (left) and a three-dimensional rendering (right).}  
	\label{T34-flat}  
\end{figure}

For $n=1$, rotation changes the energy-density distribution only
modestly, while the angular-momentum density has a pronounced toroidal
profile, as shown in Fig.~\ref{T34-flat}. The latter provides a useful
reference when interpreting the corresponding Kerr configurations.

\subsection{Kerr black holes with sphaleron clouds}

We now address the main new result: the continuation of electroweak
sphaleron configurations to rotating BH backgrounds. Switching
on $J_K$, or equivalently the horizon angular velocity $\Omega_H$,
produces stationary solutions with nonvanishing electric potentials and
matter angular momentum. We restrict the Kerr numerical analysis to
$n=1$; thus the results in this subsection concern the ordinary
sphaleron sector, not Kerr multi-sphalerons.

\begin{figure}[t!]
\setlength{\unitlength}{1cm}
\begin{picture}(15,18)
\put(-1.,0){\epsfig{file=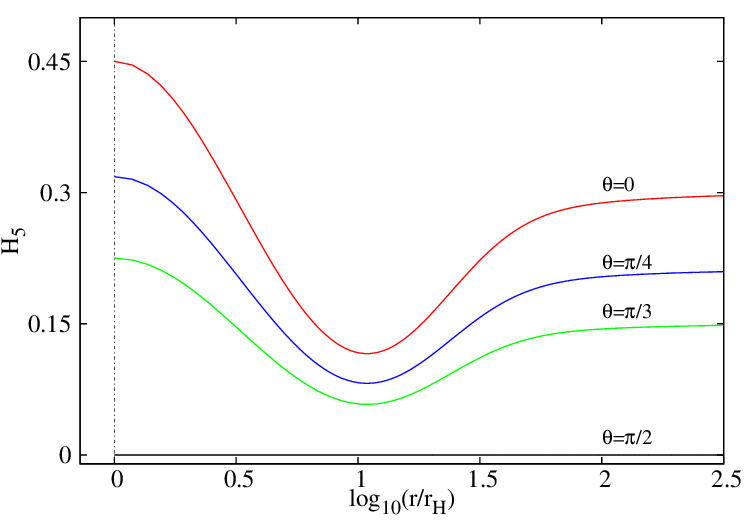,width=9cm}}
\put(8,-0.){\epsfig{file=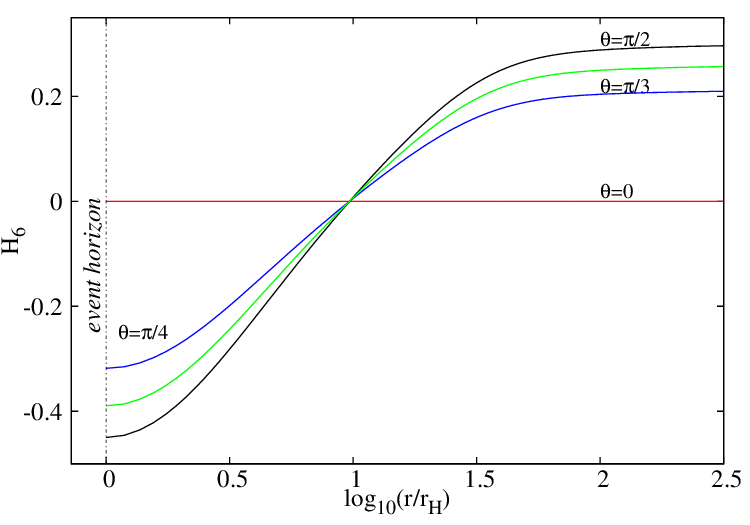,width=9cm}}
\put(-1,6){\epsfig{file=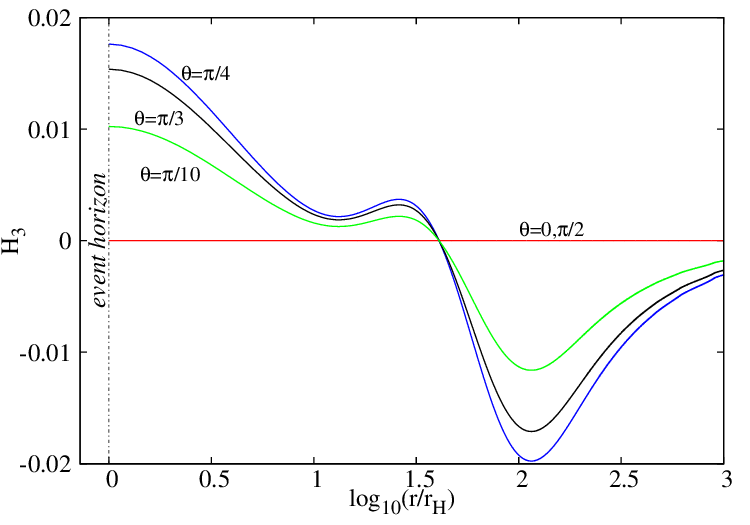,width=9cm}}
\put(8,6){\epsfig{file=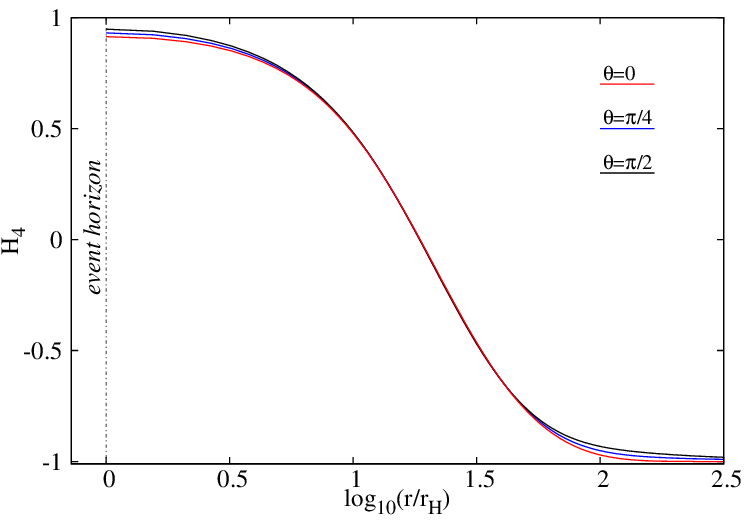,width=9cm}}
\put(-1,12){\epsfig{file=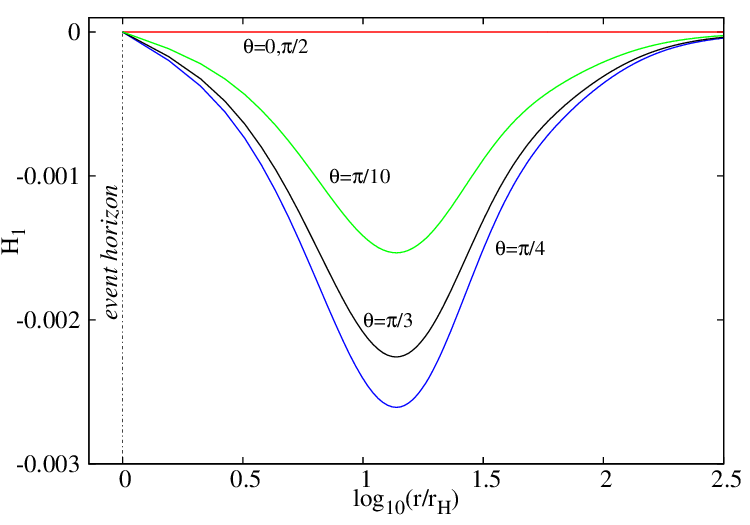,width=9cm}}
\put(8,12){\epsfig{file=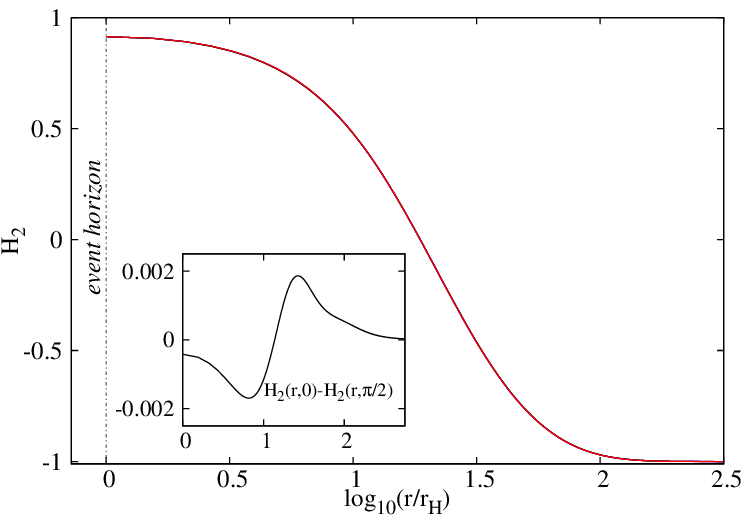,width=9cm}}
\end{picture}
    \caption{The profiles of the Yang-Mills potentials $H_i$ are shown as a function of radial coordinate and several different angles.
    Here the
 spinning sphaleron is considered in a Kerr BH background with 
 $r_H=0.09$, $\Omega_H=0.45$.  
    }
    \label{Hi-Kerr}
\end{figure}

A horizon changes the relation between the charge and the matter
angular momentum. Integrating the divergence identity (\ref{J2}) over
the exterior region now yields both an asymptotic and a horizon term.
For $n=1$ this can be written as
\begin{equation}
\label{JQrelBH}
J=\frac{4\pi}{e}{\cal Q}-J_{\rm horizon},
\end{equation}
where, with the conventions of Eq.~(\ref{J2}),
\begin{equation}
\label{JQrelBH2}
\begin{aligned}
J_{\rm horizon}={}&4\pi\int_0^{\pi/2}d\theta\,
\sqrt{-g}\,\bigg[
2\operatorname{Tr}\bigl({\cal W}F^{rt}\bigr)+
\left(A_\varphi-\frac{1}{g'}\right)f^{rt}
\bigg]_{r=r_H}.
\end{aligned}
\end{equation}
The horizon contribution breaks the simple proportionality
(\ref{JQrel}) and allows the matter to rotate even when $V=0$.
It also changes the integrated electric-sector identity, because the
horizon contributes a second boundary term to Eq.~(\ref{Me1}).

\begin{figure}[t!]
\setlength{\unitlength}{1cm}
\begin{picture}(15,18)
\put(-1.,0){\epsfig{file=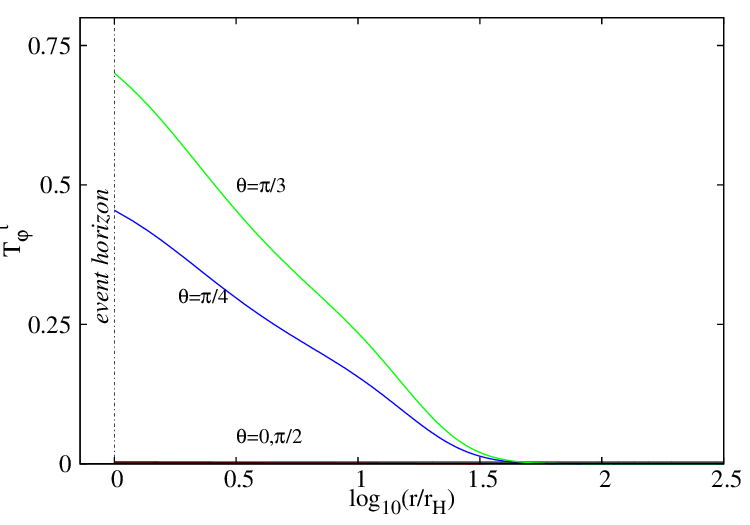,width=9cm}}
\put(8,-0.){\epsfig{file=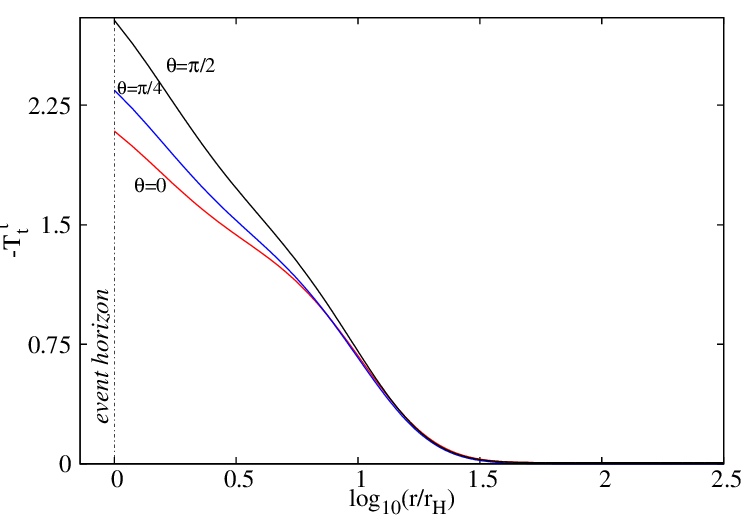,width=9cm}}
\put(-1,6){\epsfig{file=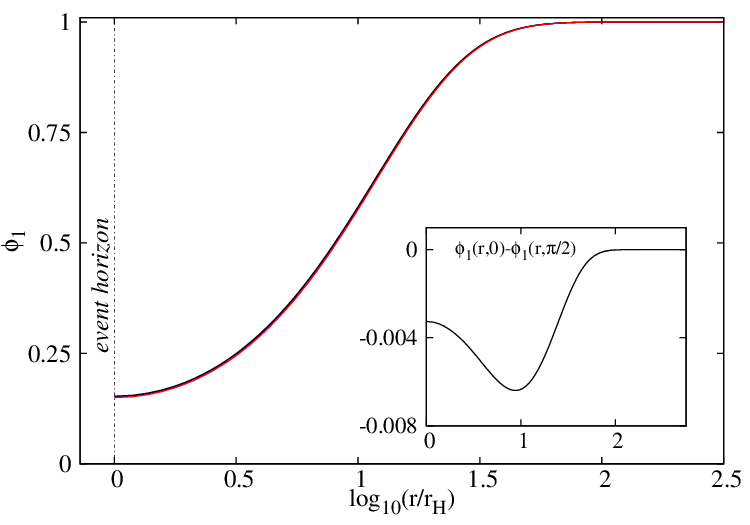,width=9cm}}
\put(8,6){\epsfig{file=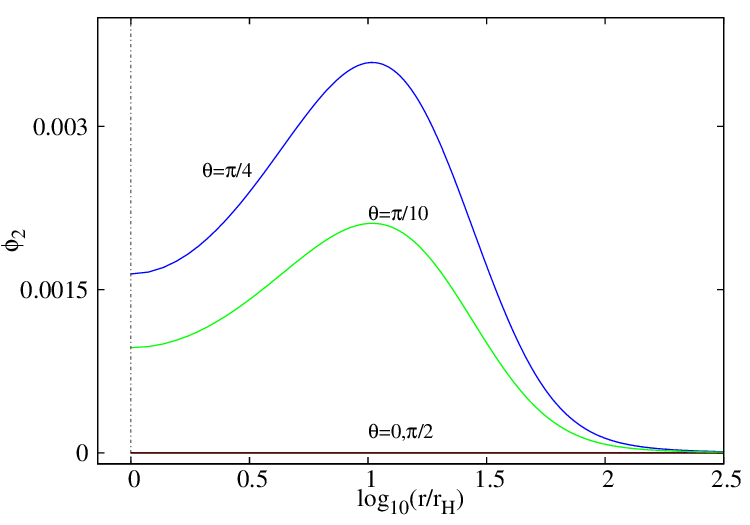,width=9cm}}
\put(-1,12){\epsfig{file=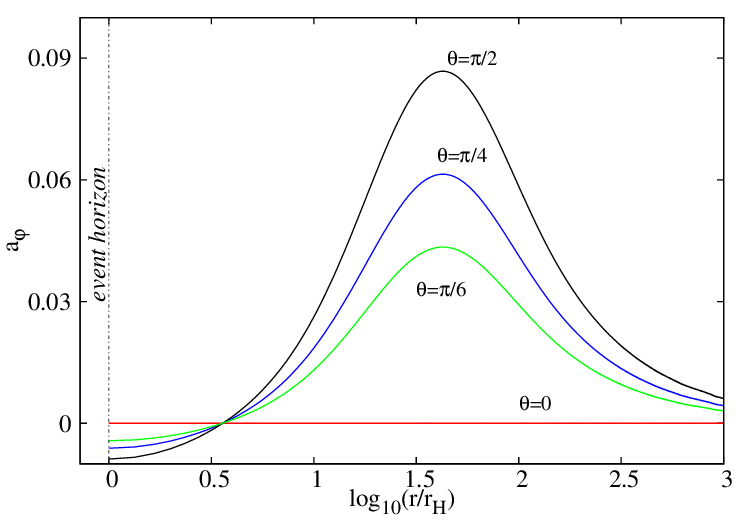,width=9cm}}
\put(8,12){\epsfig{file=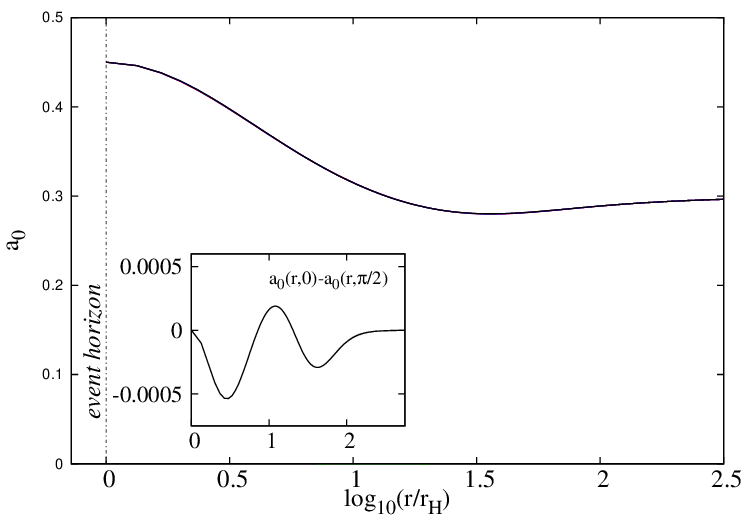,width=9cm}}
\end{picture}
    \caption{Same as Figure \ref{Hi-Kerr} for the U(1) potentials, Higgs functions and also energy and angular momentum densities.  
    }
    \label{other-Kerr}
\end{figure}

Typical field profiles are displayed in
Figs.~\ref{Hi-Kerr} and \ref{other-Kerr}. Although several gauge and
Higgs functions remain only weakly angle dependent, the matter
densities exhibit clear axial structure. In particular, the angular-
momentum density is concentrated away from the symmetry axis.

\paragraph{Branches and limiting configurations.}
We scan the solution space by fixing $\Omega_H$ and $V$ and varying
$r_H$, using the two Kerr branches described in
Section~\ref{KerrBH}. At sufficiently small $\Omega_H$, a fundamental
cloud branch joins the flat-space spinning sphaleron as the BH
size tends to zero. It reaches a turning point and continues into an
excited branch. Along the latter, the matter mass grows without bound
as $r_H\to0$, while the matter angular momentum approaches a finite
value in the cases examined. This behaviour parallels the static
Schwarzschild result in Section~\ref{Sphaleron_Schwarzschild}.

At larger $\Omega_H$, the turning-point structure also involves the
second, more rapidly spinning Kerr-background branch. The two metric
branches meet at the maximum allowed $U=\Omega_Hr_H$,
Eq.~(\ref{Umax}); this should not be confused with a merger of two
matter branches. {With the notation of} Eq.~(\ref{roots}), $x_{\rm up}$ 
connects to the small-mass/flat-space
limit at fixed $\Omega_H$, whereas $x_{\rm low}$ reaches the extremal
Kerr limit.

An extremal Kerr background is obtained by taking the appropriate
$r_H\to0$ limit with nonzero $M_K$. Since this is a distinct limit from
shrinking the BH mass to zero, results near extremality should
be interpreted separately from the flat-space limit.

\begin{figure}[t]
\centering
\includegraphics[width=.48\linewidth]{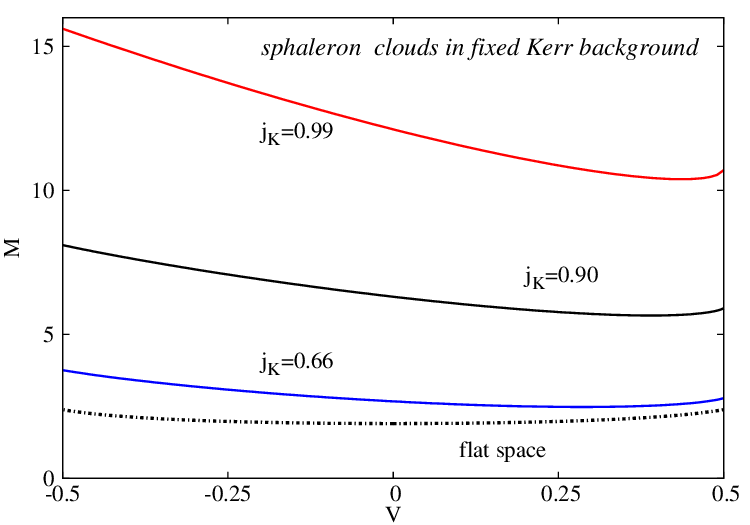}\hfill
\includegraphics[width=.48\linewidth]{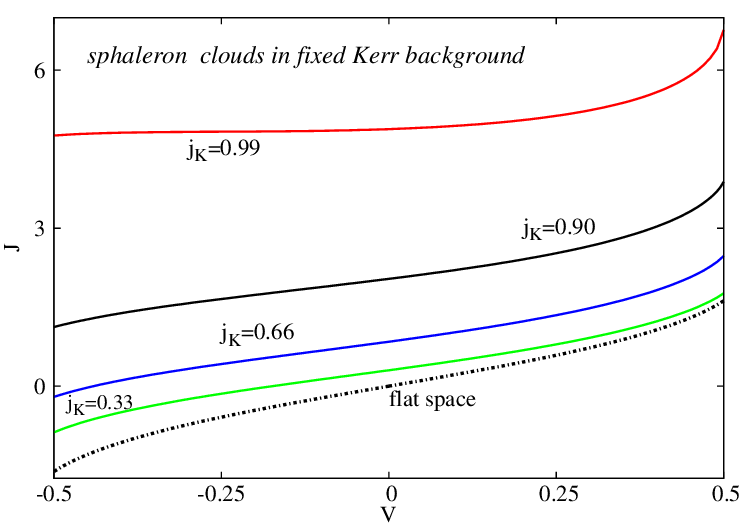}
\caption{Matter mass (left) and angular momentum (right) of Kerr
sphaleron clouds versus the asymptotic electrostatic parameter $V$,
for the indicated values of the reduced background spin
$j_K=J_K/M_K^2$. Flat-space results are included for comparison.}
\label{sph-Kerr}
\end{figure}

Figure~\ref{sph-Kerr} illustrates how rotation modifies the dependence
of the matter mass and angular momentum on $V$. Unlike the flat-space
case, the curves are not generally symmetric under $V\to -V$. In the
examples shown, $J$ increases with $V$, whereas $M$ develops a minimum
at an intermediate value. The solutions can have $J\ne0$ at $V=0$,
consistently with the horizon contribution in
Eq.~(\ref{JQrelBH}). These are properties of the displayed families,
not universal monotonicity claims for all Kerr backgrounds.

\begin{figure}[t]
\centering
\includegraphics[width=.48\linewidth]{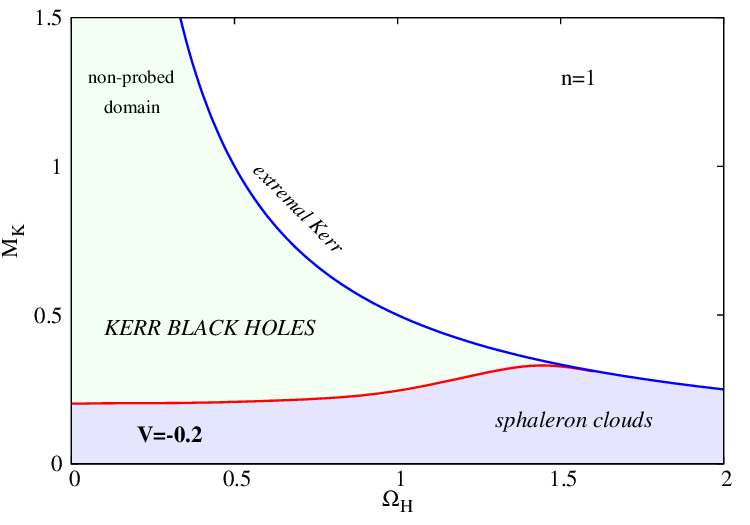}\hfill
\includegraphics[width=.48\linewidth]{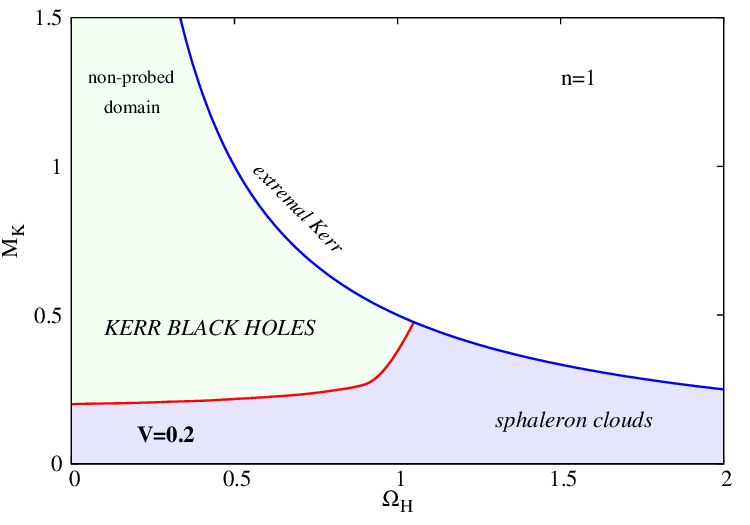}
\caption{Numerically determined domains of existence for $n=1$
sphaleron clouds in the $(M_K,\Omega_H)$ plane, for $V=-0.2$
(left) and $V=0.2$ (right). The critical boundary is inferred from
continuation of the nonlinear solutions.}
\label{domain-Kerr1}
\end{figure}

\begin{figure}[t]
\centering
\includegraphics[width=.48\linewidth]{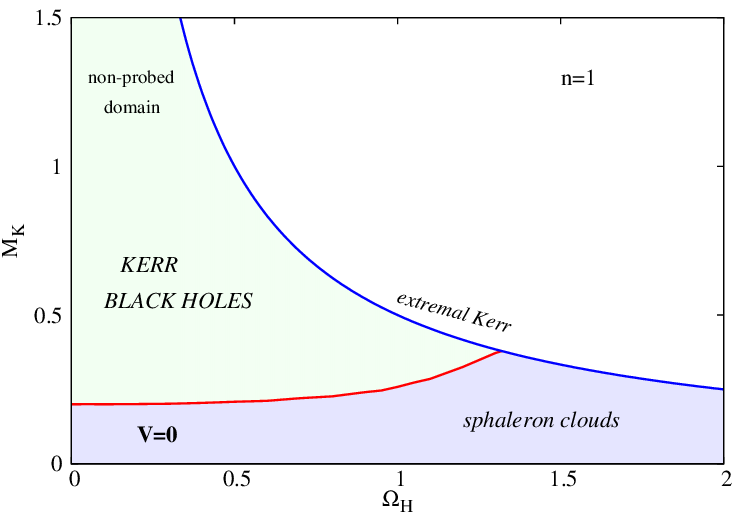}\hfill
\includegraphics[width=.48\linewidth]{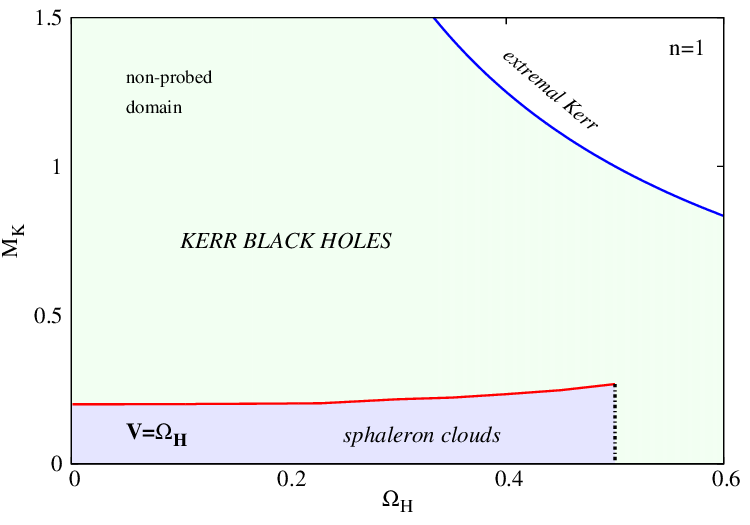}
\caption{As in Fig.~\ref{domain-Kerr1}, for $V=0$ (left) and the
one-parameter choice $V=\Omega_H$ (right). In the latter case,
the asymptotic localization bound also restricts $\Omega_H$.}
\label{domain-Kerr2}
\end{figure}

\paragraph{Domain of existence.}
For a chosen value of $V$, the clouds occupy a two-dimensional region
of the Kerr parameter space, which may be represented by
$(M_K,\Omega_H)$ in electroweak units. The numerical domains are shown
in Figs.~\ref{domain-Kerr1} and \ref{domain-Kerr2} for several
asymptotic potentials. They were inferred by continuation of many
discrete solutions and interpolation of the resulting boundary curves.
The plots identify the flat-space and Schwarzschild limits, the
extremal Kerr boundary, and a critical matter-branch curve marking the
largest background masses that support the clouds within the families
found numerically.

The numerics indicates that the critical Kerr boundary extends the
turning point found for Schwarzschild clouds. In the sampled parameter
ranges, rotation allows somewhat larger horizons than in the static
case, without a dramatic change in the characteristic sphaleron scale.
The computations also reach large horizon angular velocities when the
horizon is sufficiently small, although they become increasingly
demanding in that regime. We do not infer from this numerical trend a
rigorous absence of an upper bound on $\Omega_H$.

The special choice $V=\Omega_H$ deserves separate mention. It is a
particular parameter slice, not a synchronization condition imposed on
all solutions. Along it, the asymptotic bound $|V|\leq1/2$ also bounds
$|\Omega_H|$ in the adopted dimensionless units.

\begin{figure}[t]
\centering
\includegraphics[width=.48\linewidth]{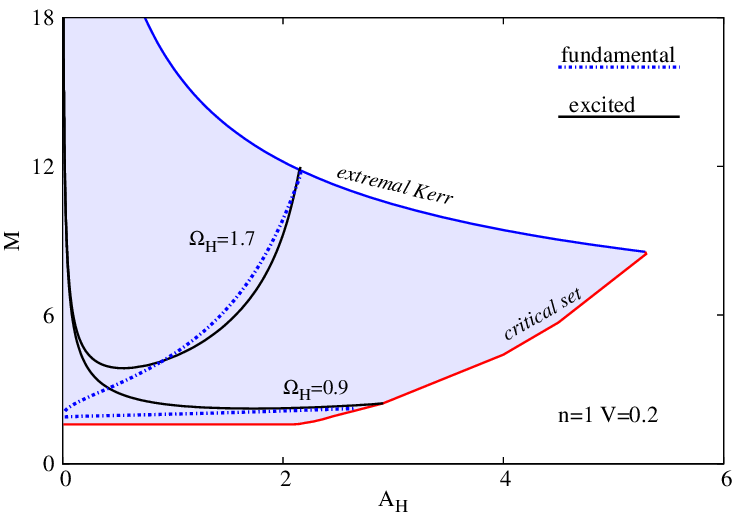}\hfill
\includegraphics[width=.48\linewidth]{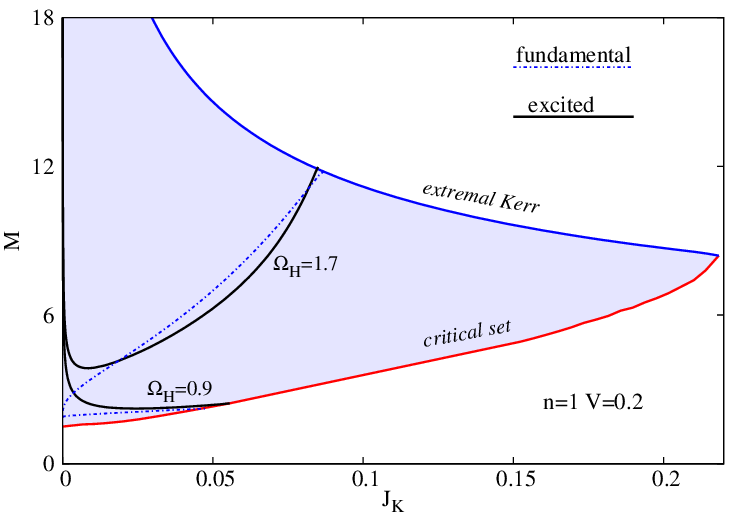}
\caption{Allowed values of the matter mass of Kerr sphaleron clouds
for $V=0.2$, shown against the horizon area $A_H$ (left) and
background angular momentum $J_K$ (right). Representative curves at
fixed $\Omega_H$ display the nonlinear branch structure.}
\label{sph-Kerr3}
\end{figure}

\begin{figure}[t]
\centering
\includegraphics[width=.48\linewidth]{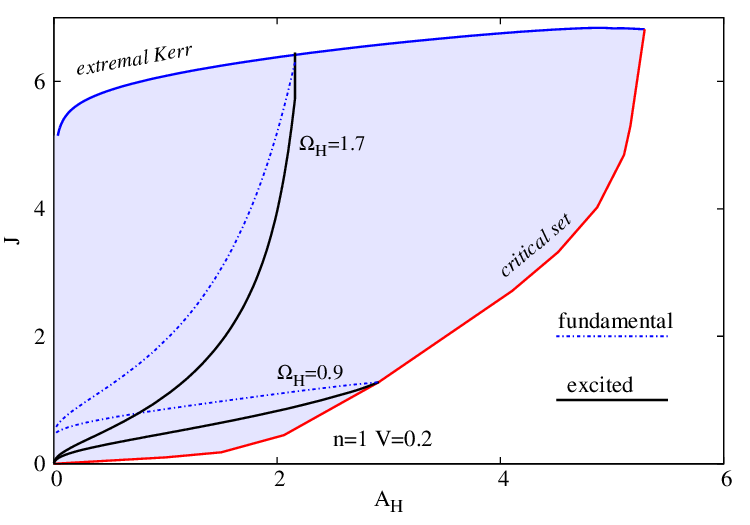}\hfill
\includegraphics[width=.48\linewidth]{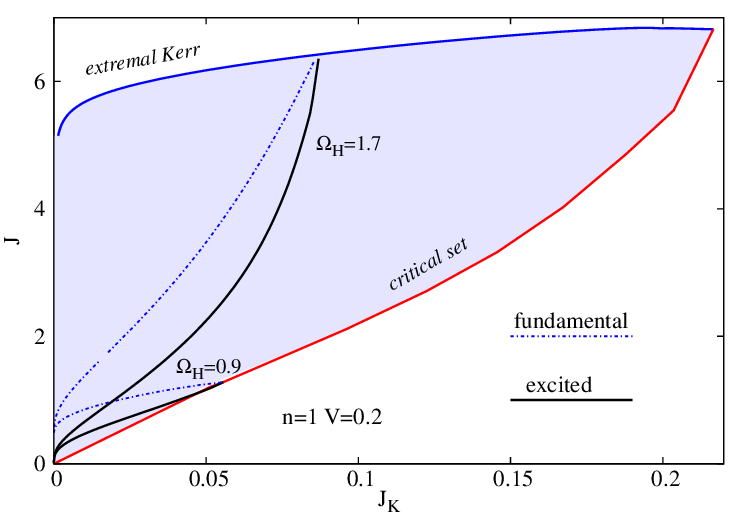}
\caption{As in Fig.~\ref{sph-Kerr3}, for the matter angular momentum
$J$ rather than its mass.}
\label{sph-Kerr4}
\end{figure}

A complementary representation is given in
Figs.~\ref{sph-Kerr3} and \ref{sph-Kerr4}, which show the matter mass
and angular momentum against $A_H$ and $J_K$ for $V=0.2$. The fixed-
$\Omega_H$ curves exhibit the same fundamental/excited branch structure
seen in the Schwarzschild limit. These are matter charges in the probe
approximation and should not be confused with the Kerr parameters
$M_K$ and $J_K$ of the prescribed background.

\paragraph{Finite mass gap and linear clouds.}
One of the motivations for studying rotating electroweak fields is the
possible connection with stationary linear clouds at superradiant
thresholds. In various bosonic models, nonlinear hairy configurations
can approach infinitesimal fields on special Kerr backgrounds; the
Proca--Higgs example of Ref.~\cite{Herdeiro:2024pmv} provides a useful
comparison. Our sphaleron-connected families display a different
behaviour. Over the Kerr backgrounds and parameter ranges explored,
their matter mass remains finite and nonzero (or grows along an
excited-branch limit), with no observed continuation to an
infinitesimal cloud.

This is evidence for a mass gap \emph{within the nonlinear families
constructed here}, not a proof that the complete bosonic electroweak
theory admits no linear clouds or superradiant modes. Establishing the
latter would require a separate analysis of the linearized field
equations and, potentially, different symmetry sectors or boundary
conditions. The finite mass gap nevertheless distinguishes the
sphaleron-cloud branches found here from the usual nonlinear families
connected continuously to linear stationary clouds.

\section{Conclusions and further remarks}
\label{Conclusions}

Kerr BHs and electroweak sphalerons are
remarkable solutions of two well-established
theories: General Relativity and the SM of particle physics, respectively.
In this work, we have investigated their interplay
by constructing nonlinear sphaleron clouds
around Schwarzschild and Kerr BHs
in the probe approximation.

Unlike many matter configurations considered
in the context of BH hair, the
electroweak sphaleron arises within an
experimentally established fundamental theory.
It is a non-perturbative, unstable configuration
of the gauge and Higgs fields, associated
with the energy barrier between topologically
distinct electroweak vacua
\cite{Manton:1983nd,Klinkhamer:1984di}.
Its existence and properties are therefore
of interest beyond the mathematical theory
of solitons and BHs.

As a first step, we have revisited static and
spinning (multi-)sphalerons in flat spacetime,
using the physical values of the electroweak
parameters, including the measured Higgs boson
mass. We have presented updated numerical
results for their mass, angular momentum,
electric charge and magnetic dipole moment,
together with useful fitting relations.
In particular, we have extended the
physical-parameter analysis of spinning
sphalerons to multi-sphaleron configurations.
These flat-space solutions provide the
reference configurations for the main
investigation of this work: the existence
and properties of sphaleron clouds in
BH backgrounds.

For Schwarzschild BHs, we have found
that sphaleron clouds exist only below a
maximal horizon size, set by the electroweak
length scale. For the fundamental sphaleron,
the corresponding BHs are necessarily
microscopic, with an areal horizon radius
of order $10^{-19}\,{\rm m}$.

The presence of a horizon also leads to a
non-trivial branch structure. Besides the
fundamental branch, which connects smoothly
to the regular flat-space sphaleron as
$r_H\to0$, there exists an excited branch
without a regular flat-space counterpart.

Along this second branch, the matter mass
grows without bound as the horizon shrinks.
We have related this behaviour to the
limiting Yang--Mills configuration in a
Schwarzschild background and, in
Appendix~\ref{ex-sol}, obtained an analytical
solution of the leading-order perturbative
Yang--Mills--Higgs problem that further
clarifies this limit.

Our main new results concern the extension
of these configurations to rotating Kerr
backgrounds. We have constructed stationary,
nonlinear electroweak sphaleron clouds and
explored their domain of existence as
functions of the BH parameters
and the electrostatic potential at infinity.
The Kerr solutions exhibit a rich branch
structure, generalizing that found in the
Schwarzschild case. The BH rotation
modifies the domain of existence and the
matter charges, while preserving the
qualitative distinction between fundamental
and excited configurations. In particular,
the latter retain the divergent-mass
behaviour in the appropriate small-horizon
limit.

An important result is that the
sphaleron-connected Kerr clouds retain
a finite mass gap: within the families
constructed here, the matter fields do
not approach an infinitesimal linear
configuration. This distinguishes them
from the familiar bosonic clouds that
occur at the threshold of superradiant
instabilities.

This finding does not, however, establish
the absence of linear electroweak clouds
or superradiant instabilities in the
full bosonic SM. Our results
concern a particular sector of stationary,
axially symmetric configurations, subject
to boundary conditions that ensure their
connection to the ordinary sphaleron.
Other families of electroweak configurations,
obeying different boundary conditions,
may admit a linear-cloud limit.
Investigating this possibility requires
a separate analysis of the corresponding
linearized field equations on Kerr.

A particularly interesting direction is
suggested by the Proca--Higgs model, for
which stationary linear clouds on Kerr
backgrounds have been reported
\cite{Herdeiro:2024pmv}.
Recent work has also established the
existence of smooth, finite-energy solitons
in the bosonic $SU(2)\times U(1)$
Weinberg--Salam theory, dubbed
\textit{electroweak balls}
\cite{Herdeiro:2026gbo};
see also Ref.~\cite{dosSantosCostaFilho:2026ach}.
These configurations provide electroweak
counterparts of the static Proca--Higgs
balls \cite{Herdeiro:2023lze}, with
the vector-boson masses generated
by the Higgs mechanism.
It would be particularly interesting
to investigate whether spinning
generalizations of electroweak balls
can support stationary clouds around
Kerr BHs, possibly admitting
a linear limit analogous to that of
the Proca--Higgs configurations
\cite{Herdeiro:2024pmv}.

Another natural extension is to include
the backreaction of the electroweak
fields on the spacetime geometry.
Although the gravitational coupling
at the electroweak scale is extremely
small, backreaction may become important
along the excited branches, where
the matter mass diverges in the
probe approximation.

Related Einstein--Yang--Mills--Higgs
systems exhibit non-trivial branch
structures and regular horizonless
limits when gravity is included
\cite{Greene:1992fw}.
It would be interesting to determine
whether the divergent excited branches
found here admit analogous regularizations
in the fully coupled
Einstein--electroweak theory.

The dynamical stability of the
BH sphaleron configurations
is another open question.
Since ordinary sphalerons are unstable,
it would be particularly relevant
to investigate how the presence
of a horizon and its rotation affect
their perturbation spectrum.

Finally, the physical interpretation
of these results deserves emphasis.
The Kerr geometry provides a realistic
description of the exterior of
astrophysical rotating BHs,
while the sphaleron is a
non-perturbative solution of the
electroweak sector of the Standard
Model. Although BHs with
horizon sizes comparable to the
electroweak length scale remain
hypothetical, their study brings
together two physically well-founded
ingredients in a controlled
theoretical setting.

The solutions constructed here
demonstrate that nonlinear
Standard-Model gauge--Higgs
configurations can coexist with
Schwarzschild and Kerr horizons,
providing a starting point for
further investigations of
non-perturbative electroweak
phenomena in BH spacetimes.

\section*{Acknowledgement}
 This work is supported by the Center for Research and Development in Mathematics and Applications (CIDMA) (\url{https://ror.org/05pm2mw36}) under the Portuguese Foundation for Science and Technology 
(FCT -- Fundaç\~ao para a Ci\^encia e a Tecnologia, \url{https://ror.org/00snfqn58}), Grants UID/04106/2025 (\url{https://doi.org/10.54499/UID/04106/2025}) and UID/PRR/04106/2025 (\url{https://doi.org/10.54499/UID/PRR/04106/2025}), as well as the projects: Horizon Europe staff exchange (SE) programme HORIZON-MSCA2021-SE-01 Grant No.\ NewFunFiCO-101086251 and 2022.04560.PTDC (\url{https://doi.org/10.54499/2022.04560.PTDC}).
E. Radu also gratefully acknowledges the support of the Alexander von Humboldt Foundation.

\appendix
 
\section{Field-strength and Higgs covariant derivatives components}
\label{expressions}

For completeness, we collect the explicit field strengths and Higgs
covariant derivatives corresponding to the Minkowski/Schwarzschild
ansatz~(\ref{YM-ansatz})--(\ref{H-ansatz}). We write
$F_{\mu\nu}=\frac12 F_{\mu\nu}^{(a)}\tau_a^{(n)}$,
where the parenthesized labels refer to the rotating isospin basis,
and use $H_{i,r}=\partial_r H_i$ and
$H_{i,\theta}=\partial_\theta H_i$.
Antisymmetric components not displayed below follow from
$F_{\nu\mu}=-F_{\mu\nu}$; the same applies to $f_{\mu\nu}$.

The non-vanishing $SU(2)$ field-strength components are
\begin{align}
F_{r\varphi}^{(r)}
 &= -\frac{n\sin\theta}{gr}
       \bigl(rH_{3,r}-H_1H_4\bigr), \notag\\
F_{r\varphi}^{(\theta)}
 &= \frac{n\sin\theta}{gr}
       \bigl(rH_{4,r}+H_1H_3+H_1\cot\theta\bigr), \notag\\
F_{r\theta}^{(\varphi)}
 &= -\frac{1}{gr}
       \bigl(rH_{2,r}+H_{1,\theta}\bigr), \notag\\
F_{rt}^{(r)}
 &= \frac{1}{g}\left(H_{5,r}+\frac{H_1H_6}{r}\right),
 \label{eq:Frtr}\\
F_{rt}^{(\theta)}
 &= \frac{1}{g}\left(H_{6,r}-\frac{H_1H_5}{r}\right),\notag\\
F_{\theta\varphi}^{(r)}
 &= \frac{n\sin\theta}{g}
       \bigl(1-H_2H_4-H_{3,\theta}-H_3\cot\theta\bigr),\notag\\
F_{\theta\varphi}^{(\theta)}
 &= \frac{n\sin\theta}{g}
       \bigl(H_{4,\theta}-H_2H_3
               +(H_4-H_2)\cot\theta\bigr),\notag\\
F_{\theta t}^{(r)}
 &= \frac{1}{g}\bigl(H_{5,\theta}-H_2H_6\bigr),\notag\\
F_{\theta t}^{(\theta)}
 &= \frac{1}{g}\bigl(H_{6,\theta}+H_2H_5\bigr),\notag\\
F_{\varphi t}^{(\varphi)}
 &= \frac{n\sin\theta}{g}
       \bigl(H_6\cot\theta+H_3H_6+H_4H_5\bigr).
 \notag
\end{align}

For the Abelian ansatz~(\ref{M-ansatz}), the non-vanishing components are
\begin{align}
f_{rt}&=\frac{a_{0,r}}{g'},
&f_{\theta t}&=\frac{a_{0,\theta}}{g'},\notag\\
f_{r\varphi}&=\frac{\sin\theta}{g'}a_{\varphi,r},
&f_{\theta\varphi}&=\frac{1}{g'}
  \bigl(\sin\theta\,a_{\varphi,\theta}
                   +a_\varphi\cos\theta\bigr).
\end{align}

The Higgs covariant derivatives are, in the two-component
representation of the doublet,
\begin{align}
D_r\Phi
&=\frac{iv}{2\sqrt{2}\,r}
\begin{pmatrix}
e^{-in\varphi}\bigl[
 H_1(-\cos\theta\,\Phi_1+\sin\theta\,\Phi_2)
 +2r(\sin\theta\,\Phi_{1,r}+\cos\theta\,\Phi_{2,r})
 \bigr]\\
-H_1(\sin\theta\,\Phi_1+\cos\theta\,\Phi_2)
-2r(\cos\theta\,\Phi_{1,r}-\sin\theta\,\Phi_{2,r})
\end{pmatrix},\notag\\[0.7em]
D_\theta\Phi
&=\frac{iv}{2\sqrt{2}}
\begin{pmatrix}
e^{-in\varphi}\bigl[
 (1+H_2)(\cos\theta\,\Phi_1-\sin\theta\,\Phi_2)
 +2(\sin\theta\,\Phi_{1,\theta}+\cos\theta\,\Phi_{2,\theta})
 \bigr]\\
(1+H_2)(\sin\theta\,\Phi_1+\cos\theta\,\Phi_2)
-2(\cos\theta\,\Phi_{1,\theta}-\sin\theta\,\Phi_{2,\theta})
\end{pmatrix},\notag\\[0.7em]
D_\varphi\Phi
&=\frac{v}{2\sqrt{2}}
\begin{pmatrix}
e^{-in\varphi}\bigl[
 (n(1+H_4)-a_\varphi\sin\theta)\sin\theta\,\Phi_1
 +((2n-a_\varphi\sin\theta)\cos\theta
    +nH_3\sin\theta)\Phi_2\bigr]\\
\sin\theta\bigl[
 (nH_3+a_\varphi\cos\theta)\Phi_1
 +(n(1-H_4)-a_\varphi\sin\theta)\Phi_2\bigr]
\end{pmatrix},\notag\\[0.7em]
D_t\Phi
&=\frac{v}{2\sqrt{2}}
\begin{pmatrix}
e^{-in\varphi}\bigl[
 (H_6-a_0\sin\theta)\Phi_1
 -(H_5+a_0\cos\theta)\Phi_2\bigr]\\
-\bigl[
 (H_5-a_0\cos\theta)\Phi_1
 +(H_6+a_0\sin\theta)\Phi_2\bigr]
\end{pmatrix}.
\end{align}

\section{A closed-form Higgs solution on a Schwarzschild background}
\label{ex-sol}

In this appendix we obtain a closed-form solution for the leading-order
Higgs profile on the exact, static and spherically symmetric Yang--Mills
configuration in a Schwarzschild background found in
Ref.~\cite{Boutaleb-Joutei:1979hdt} and discussed in
Section~\ref{Sphaleron_Schwarzschild}. The construction describes the
small-$\tilde\alpha$ limit of the excited branch. It is perturbative in
the Yang--Mills--Higgs system: the gauge field is held at its
Yang--Mills background value when solving for the leading Higgs profile.

Here $r_h$ denotes the \emph{areal} Schwarzschild horizon radius, not the
quasi-isotropic radius $r_H$ used in the main text. Thus
$r_h=4r_H$ for Schwarzschild. Set $y=R/r_h$, where $R$ is the areal
radial coordinate. The metric takes the form
\begin{equation}
\label{Sch1}
ds^2=-N(y)\,dt^2+r_h^2
\left[\frac{dy^2}{N(y)}+y^2
   (d\theta^2+\sin^2\theta\,d\varphi^2)\right],
\qquad N(y)=1-\frac1y.
\end{equation}
The event horizon lies at $y=1$, and spatial infinity at $y\to\infty$.
We do not use the rescalings of Section~\ref{quant} in this appendix.

Setting the Higgs self-coupling to zero, $\lambda=0$, the radial
\emph{static energy functional}, up to the overall angular factor
$4\pi$, is obtained from the integrand
\begin{equation}
\mathcal E_{\rm red}=
\frac{1}{r_h^2}\left[
 N(y)\left(\frac{dw}{dy}\right)^2
 +\frac{(1-w^2)^2}{2y^2}\right]
+\frac{v^2}{2}\left[
 y^2N(y)\left(\frac{d\phi}{dy}\right)^2
 +\frac12\phi^2(1+w)^2\right].
\end{equation}
Varying this functional yields
\begin{align}
\frac{d}{dy}\left(N\frac{dw}{dy}\right)
+\frac{w(1-w^2)}{y^2}
-\frac{\tilde\alpha^2}{4}\phi^2(1+w)&=0,
\label{neqw}\\
\frac{d}{dy}\left(y^2N\frac{d\phi}{dy}\right)
-\frac12\phi(1+w)^2&=0,
\label{neqh}
\end{align}
where
\begin{equation}
\tilde\alpha=vr_h.
\end{equation}
The parameter $\tilde\alpha$ should not be confused with the
dimensionless gravitational coupling $\alpha$ defined in
Section~\ref{backreaction}.

For $\tilde\alpha=0$, the gauge equation admits the exact solution
\cite{Boutaleb-Joutei:1979hdt}
\begin{equation}
\label{Chakrabarti}
w_0(y)=\frac{-2y+3+\sqrt3}{2y+3(1+\sqrt3)}.
\end{equation}
It satisfies $w_0(1)=2-\sqrt3$ and tends to $-1$ as $y\to\infty$.
At leading order in $\tilde\alpha$, we solve the Higgs equation
(\ref{neqh}) with $w=w_0$, while corrections to the gauge profile
enter at order $\tilde\alpha^2$.

To reduce the Higgs equation to hypergeometric form, introduce
\begin{equation}
x=\log\left(1-\frac1y\right),\qquad
 y=\frac1{1-e^x},\qquad -\infty<x<0.
\end{equation}
\begin{figure}[ht]
	\centering
	\includegraphics[width=0.65\linewidth]{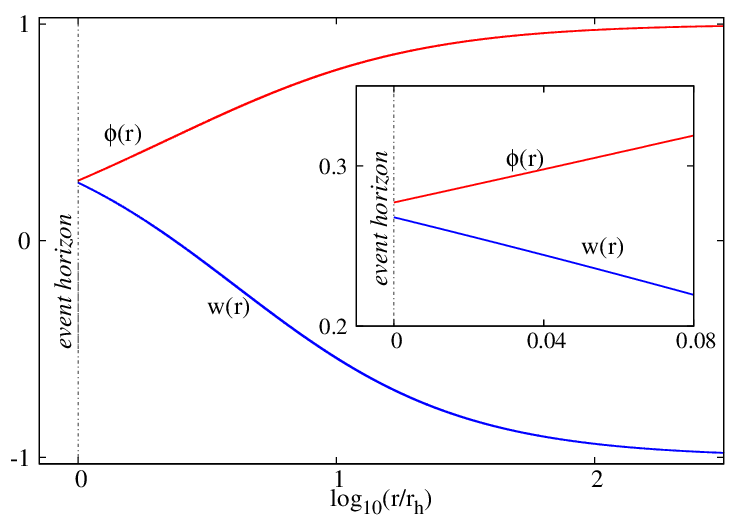}
	\caption{The leading-order Higgs profile on the exact Schwarzschild
		Yang--Mills background, Eqs.~(\ref{Chakrabarti}) and
		(\ref{ex-sol-Z}), plotted against the areal-radius ratio $R/r_h$.
		The inset resolves the near-horizon behaviour.}
	\label{sphaleron-BH-sol}
\end{figure}
The horizon corresponds to $x\to-\infty$, spatial infinity to
$x\to0^-$, and $N=e^x$. With this change of variable, the equation
can be solved in terms of Gauss hypergeometric functions.

For compactness, define
\begin{equation}
P(R)=2+3(1+\sqrt3)\frac{r_h}{R},\qquad
z(R)=1+3(-2+\sqrt3)\left(1-\frac{r_h}{R}\right),
\end{equation}
and $a_\pm=(1\pm\sqrt5)/2$. Returning to the areal radius $R$,
the general leading-order Higgs solution is
\begin{align}
\label{ex-sol-Z}
\phi(R)=P(R)^{a_-}\Big[&
 s_1\,{}_2F_1(a_-,a_-;2a_-;z(R))\\
 &+s_2\,P(R)^{\sqrt5}
 {}_2F_1(a_+,a_+;2a_+;z(R))\Big].\notag
\end{align}
Here ${}_2F_1$ is the Gauss hypergeometric function. The integration
constants $s_1$ and $s_2$ are fixed by regularity at the horizon and
by $\phi(R\to\infty)=1$. In this convention, the regularity condition
requires the two logarithmic terms at $z\to1$ to cancel.

For the solution written above, the resulting approximate coefficients
are
\begin{equation}
s_1\simeq1.63124,\qquad s_2\simeq-0.000845689.
\end{equation}
They give the finite horizon value
$\phi(r_h)\simeq0.27719$ and the asymptotic behaviour
\begin{equation}
\phi(R)=1-2.78551\frac{r_h}{R}
 +O\!\left(\frac{r_h^2}{R^2}\right).
\end{equation}

The total matter energy admits the perturbative expansion
\begin{equation}
\label{MtotYMH}
M=-4\pi\int_{r_h}^{\infty}dR\,R^2T_t{}^t
  =M_{\rm YM}+\tilde\alpha^2 {M_{\rm Higgs}^{(2)} }
    +O\!\left(\frac{\tilde\alpha^4}{r_h}\right),
\end{equation}
where the two reported leading coefficients are
\begin{align}
\label{MtotYM}
M_{\rm YM}&=\frac{4\pi}{r_h}
 (0.47949013476947977\ldots),\\
\label{MtotH}
{M_{\rm Higgs}^{(2)}}&=\frac{4\pi}{r_h}
 (1.3927567317183576\ldots).
\end{align} 
For $\tilde\alpha\lesssim10^{-3}$, we find close agreement between the numerical excited-branch
solutions and the leading-order analytical expressions.

 \bibliographystyle{unsrt}

\bibliography{biblio}

\end{document}